\pdfoutput=1
\documentclass[12pt,a4paper]{article}

\usepackage{ifthen} 
\newboolean{pdflatex}
\setboolean{pdflatex}{true} 

\newboolean{articletitles}
\setboolean{articletitles}{true} 

\newboolean{uprightparticles}
\setboolean{uprightparticles}{false} 

\newboolean{inbibliography}
\setboolean{inbibliography}{false} 

\usepackage{microtype}
\usepackage{lineno}  
\usepackage{xspace} 

\usepackage{graphicx}  
\usepackage{color}
\usepackage{colortbl}
\graphicspath{{./figs/}} 

\usepackage{amsmath} 
\usepackage{amssymb}
\usepackage{amsfonts}
\usepackage{upgreek} 

\newcommand*\patchAmsMathEnvironmentForLineno[1]{%
\expandafter\let\csname old#1\expandafter\endcsname\csname #1\endcsname
\expandafter\let\csname oldend#1\expandafter\endcsname\csname
end#1\endcsname
 \renewenvironment{#1}%
   {\linenomath\csname old#1\endcsname}%
   {\csname oldend#1\endcsname\endlinenomath}%
}
\newcommand*\patchBothAmsMathEnvironmentsForLineno[1]{%
  \patchAmsMathEnvironmentForLineno{#1}%
  \patchAmsMathEnvironmentForLineno{#1*}%
}
\AtBeginDocument{%
\patchBothAmsMathEnvironmentsForLineno{equation}%
\patchBothAmsMathEnvironmentsForLineno{align}%
\patchBothAmsMathEnvironmentsForLineno{flalign}%
\patchBothAmsMathEnvironmentsForLineno{alignat}%
\patchBothAmsMathEnvironmentsForLineno{gather}%
\patchBothAmsMathEnvironmentsForLineno{multline}%
}

\usepackage{hyperref}    
\usepackage[all]{hypcap}

\def\lhcb {\mbox{LHCb}\xspace}

\def\rich   {RICH\xspace}

\ifthenelse{\boolean{uprightparticles}}%
{

 \def\Peta        {\ensuremath{\upeta}\xspace}

 \def\Ppi         {\ensuremath{\uppi}\xspace}

 \def\PDelta      {\ensuremath{\Delta}\xspace}                 
 \def\PXi      {\ensuremath{\Xi}\xspace}                 
 \def\PLambda      {\ensuremath{\Lambda}\xspace}                 
 \def\PSigma      {\ensuremath{\Sigma}\xspace}                 
 \def\POmega      {\ensuremath{\Omega}\xspace}                 
 \def\PUpsilon      {\ensuremath{\Upsilon}\xspace}

 \def\PB      {\ensuremath{\mathrm{B}}\xspace}                 
                  
 \def\PD      {\ensuremath{\mathrm{D}}\xspace}

 \def\PK      {\ensuremath{\mathrm{K}}\xspace}

 \def\Pb      {\ensuremath{\mathrm{b}}\xspace}                 
 \def\Pc      {\ensuremath{\mathrm{c}}\xspace}                 
 \def\Pd      {\ensuremath{\mathrm{d}}\xspace}

 \def\Ph      {\ensuremath{\mathrm{h}}\xspace}                 
 \def\Pi      {\ensuremath{\mathrm{i}}\xspace}

 \def\Pp      {\ensuremath{\mathrm{p}}\xspace}

 \def\Ps      {\ensuremath{\mathrm{s}}\xspace}

}
{

 \def\Peta        {\ensuremath{\eta}\xspace}

 \def\Ppi         {\ensuremath{\pi}\xspace}

 \mathchardef\PDelta="7101
 \mathchardef\PXi="7104
 \mathchardef\PLambda="7103
 \mathchardef\PSigma="7106
 \mathchardef\POmega="710A
 \mathchardef\PUpsilon="7107
                  
 \def\PB      {\ensuremath{B}\xspace}                 
                  
 \def\PD      {\ensuremath{D}\xspace}

 \def\PK      {\ensuremath{K}\xspace}

 \def\Pb      {\ensuremath{b}\xspace}                 
 \def\Pc      {\ensuremath{c}\xspace}                 
 \def\Pd      {\ensuremath{d}\xspace}

 \def\Ph      {\ensuremath{h}\xspace}                 
 \def\Pi      {\ensuremath{i}\xspace}

 \def\Pp      {\ensuremath{p}\xspace}

 \def\Ps      {\ensuremath{s}\xspace}

}

\def\dquark    {\ensuremath{\Pd}\xspace}

\def\squark    {\ensuremath{\Ps}\xspace}

\def\cquark    {\ensuremath{\Pc}\xspace}

\def\bquark    {\ensuremath{\Pb}\xspace}

\def\pion  {\ensuremath{\Ppi}\xspace}
\def\piz   {\ensuremath{\pion^0}\xspace}

\def\pip   {\ensuremath{\pion^+}\xspace}
\def\pim   {\ensuremath{\pion^-}\xspace}

\def\pimp  {\ensuremath{\pion^\mp}\xspace}

\def\kaon  {\ensuremath{\PK}\xspace}
  \def\Kbar  {\kern 0.2em\overline{\kern -0.2em \PK}{}\xspace}

\def\Kz    {\ensuremath{\kaon^0}\xspace}
\def\Kzb   {\ensuremath{\Kbar^0}\xspace}

\def\Km    {\ensuremath{\kaon^-}\xspace}
\def\Kpm   {\ensuremath{\kaon^\pm}\xspace}

\def\KS    {\ensuremath{\kaon^0_{\rm\scriptscriptstyle S}}\xspace} 
 
\def\Kstarz  {\ensuremath{\kaon^{*0}}\xspace}

\def\Kstarp  {\ensuremath{\kaon^{*+}}\xspace}

\newcommand{\etapr}{\ensuremath{\Peta^{\prime}}\xspace}

  \def\Dbar    {\kern 0.2em\overline{\kern -0.2em \PD}{}\xspace}
\def\D       {\ensuremath{\PD}\xspace}

\def\Dz      {\ensuremath{\D^0}\xspace}

\def\Dsm     {\ensuremath{\D^-_\squark}\xspace}

\def\Dsmp    {\ensuremath{\D^{\mp}_\squark}\xspace}

\def\B       {\ensuremath{\PB}\xspace}
\def\Bbar    {\ensuremath{\kern 0.18em\overline{\kern -0.18em \PB}{}}\xspace}

\def\Bz      {\ensuremath{\B^0}\xspace}

\def\Bu      {\ensuremath{\B^+}\xspace}
\def\Bub     {\ensuremath{\B^-}\xspace}
\def\Bp      {\ensuremath{\Bu}\xspace}

\def\Bd      {\ensuremath{\B^0}\xspace}
\def\Bs      {\ensuremath{\B^0_\squark}\xspace}

  \def\Y#1S{\ensuremath{\PUpsilon{(#1S)}}\xspace}

\def\proton      {\ensuremath{\Pp}\xspace}

\def\Lz {\ensuremath{\PLambda}\xspace}
\def\Lbar {\ensuremath{\kern 0.1em\overline{\kern -0.1em\PLambda}}\xspace}

\def\Lb      {\ensuremath{\Lz^0_\bquark}\xspace}
\def\Lbbar   {\ensuremath{\bar{\Lz}^0_\bquark}\xspace}
\def\Xib     {\ensuremath{\PXi^{0}_{\bquark}}\xspace}
\def\LbXib   {\ensuremath{\Lb (\Xib)}\xspace}
\def\Lc      {\ensuremath{\Lz^+_\cquark}\xspace}

\newcommand{\decay}[2]{\ensuremath{#1\!\to #2}\xspace}         

\def\to                 {\ensuremath{\rightarrow}\xspace}

\def\CP                {\ensuremath{C\!P}\xspace}

\newcommand{\ACP}{\ensuremath{{\cal A}^{\CP}}\xspace}

\def\LL   {Long\xspace}
\def\DD   {Downstream\xspace}

\def\KSph         {\ensuremath{\KS\proton\Ph^-}\xspace}
\def\KSppi        {\ensuremath{\KS\proton\pim}\xspace}
\def\KSpK         {\ensuremath{\KS\proton\Km}\xspace}
\def\KSpipi       {\ensuremath{\KS\pip\pim}\xspace}

\def\LbXibToKSph   {\decay{\LbXib}{\KSph}}
\def\LbXibToKSppi  {\decay{\LbXib}{\KSppi}}
\def\LbXibToKSpK   {\decay{\LbXib}{\KSpK}}

\def\LbToKSph     {\decay{\Lb}{\KSph}}
\def\LbToKSppi    {\decay{\Lb}{\KSppi}}
\def\LbToKSpK     {\decay{\Lb}{\KSpK}}

\def\LbToKzbppi   {\decay{\Lb}{\Kzb\proton\pim}}
\def\LbToKzpK     {\decay{\Lb}{\Kz \proton\Km}}

\def\XibToKSppi   {\decay{\Xib}{\KSppi}}
\def\XibToKSpK    {\decay{\Xib}{\KSpK}}

\def\XibToKzbppi  {\decay{\Xib}{\Kzb\proton\pim}}
\def\XibToKzbpK   {\decay{\Xib}{\Kzb\proton\Km}}

\def\LbToLcpi        {\decay{\Lb}{\Lc \pim}}
\def\LbToLcpiTopKS   {\decay{\Lb}{\Lc (\to \proton \KS) \pim}}
\def\LbToLcpiTopKz   {\decay{\Lb}{\Lc (\to \proton \Kzb) \pim}}

\def\LbToLcK         {\decay{\Lb}{\Lc \Km}}
\def\LbToLcKTopKS    {\decay{\Lb}{\Lc (\to \proton \KS) \Km}}
\def\LbToLcKTopKz    {\decay{\Lb}{\Lc (\to \proton \Kzb) \Km}}
\def\LbToLcKToppiK   {\decay{\Lb}{\Lc (\to \proton \Km \pip) \Km}}

\def\LbToLchTopKS    {\decay{\Lb}{\Lc (\to \proton \KS) \Ph^{-}}}

\def\LbToDsp   {\decay{\Lb}{\Dsm \proton}}
\def\LbToDspToKsK   {\decay{\Lb}{\Dsm (\to \KS \Km) \proton}}
\def\LbToDspToKzK   {\decay{\Lb}{\Dsm (\to \Kz \Km) \proton}}

\def\LbToLchToppiK {\decay{\Lb}{\Lc (\to \proton \Km\pip) h^{-}}}

\def\BToKSpipi  {\decay{\B}{\KS \pip \pim}}

\def\BdToKSpipi  {\decay{\Bd}{\KS \pip \pim}}
\def\BdToKzpipi  {\decay{\Bd}{\Kz \pip \pim}}
\def\BsToKSpipi  {\decay{\Bs}{\KS \pip \pim}}

\def\BsToKSKpi   {\decay{\Bs}{\KS \Kpm \pimp}}

\def\BdToetapKs {\decay{\Bd}{\etapr (\rho^{0} \gamma) \KS}}
\def\BdToKsPiPiGamma   {\decay{\Bd}{\KS \pip \pim \gamma}}
\def\BdToKstrho {\decay{\Bd}{\Kstarz (\to \KS \piz) \rho^{0}}}
\def\BuToKstrho {\decay{\Bu}{\Kstarp (\to \KS \pip) \rho^{0}}}
\def\BuToDzpi {\decay {\Bub} {\Dz (\to \KS \pip \pim) \pim}}

\def\AT#1     {\ensuremath{A_{\mathrm{T}}^{#1}}\xspace}           

\def\C#1      {\ensuremath{\mathcal{C}_{#1}}\xspace}                       
\def\Cp#1     {\ensuremath{\mathcal{C}_{#1}^{'}}\xspace}                    
\def\Ceff#1   {\ensuremath{\mathcal{C}_{#1}^{\mathrm{(eff)}}}\xspace}        
\def\Cpeff#1  {\ensuremath{\mathcal{C}_{#1}^{'\mathrm{(eff)}}}\xspace}       
\def\Ope#1    {\ensuremath{\mathcal{O}_{#1}}\xspace}                       
\def\Opep#1   {\ensuremath{\mathcal{O}_{#1}^{'}}\xspace}                    

\newcommand{\tev}{\ifthenelse{\boolean{inbibliography}}{\ensuremath{~T\kern -0.05em eV}\xspace}{\ensuremath{\mathrm{\,Te\kern -0.1em V}}\xspace}}
\newcommand{\gev}{\ensuremath{\mathrm{\,Ge\kern -0.1em V}}\xspace}
\newcommand{\mev}{\ensuremath{\mathrm{\,Me\kern -0.1em V}}\xspace}
\newcommand{\kev}{\ensuremath{\mathrm{\,ke\kern -0.1em V}}\xspace}
\newcommand{\ev}{\ensuremath{\mathrm{\,e\kern -0.1em V}}\xspace}
\newcommand{\gevc}{\ensuremath{{\mathrm{\,Ge\kern -0.1em V\!/}c}}\xspace}
\newcommand{\mevc}{\ensuremath{{\mathrm{\,Me\kern -0.1em V\!/}c}}\xspace}
\newcommand{\gevcc}{\ensuremath{{\mathrm{\,Ge\kern -0.1em V\!/}c^2}}\xspace}
\newcommand{\gevgevcccc}{\ensuremath{{\mathrm{\,Ge\kern -0.1em V^2\!/}c^4}}\xspace}
\newcommand{\mevcc}{\ensuremath{{\mathrm{\,Me\kern -0.1em V\!/}c^2}}\xspace}

\def\mm   {\ensuremath{\rm \,mm}\xspace}

\def\mum  {\ensuremath{{\,\upmu\rm m}}\xspace}

\def\invfb   {\ensuremath{\mbox{\,fb}^{-1}}\xspace}

\def\ps   {\ensuremath{{\rm \,ps}}\xspace}

\newcommand{\stat}{\ensuremath{\mathrm{\,(stat)}}\xspace}
\newcommand{\syst}{\ensuremath{\mathrm{\,(syst)}}\xspace}

\newcommand{\chisq}{\ensuremath{\chi^2}\xspace}

\newcommand{\chisqip}{\ensuremath{\chi^2_{\rm IP}}\xspace}
\newcommand{\chisqvs}{\ensuremath{\chi^2_{\rm VS}}\xspace}
\newcommand{\chisqvtx}{\ensuremath{\chi^2_{\rm vtx}}\xspace}

\def\gsim{{~\raise.15em\hbox{$>$}\kern-.85em
          \lower.35em\hbox{$\sim$}~}\xspace}
\def\lsim{{~\raise.15em\hbox{$<$}\kern-.85em
          \lower.35em\hbox{$\sim$}~}\xspace}

\def\sPlot{\mbox{\em sPlot}}

\def\pt         {\mbox{$p_{\rm T}$}\xspace}
\def\et         {\mbox{$E_{\rm T}$}\xspace}

\def\evtgen     {\mbox{\textsc{EvtGen}}\xspace}

\def\geant      {\mbox{\textsc{Geant4}}\xspace}

\def\photos     {\mbox{\textsc{Photos}}\xspace}

\def\pythia     {\mbox{\textsc{Pythia}}\xspace}

\def\tell1  {TELL1\xspace}
\def\ukl1   {UKL1\xspace}

\newcommand{\eg}{\mbox{\itshape e.g.}\xspace}

\newcommand{\cf}{\mbox{\itshape cf.}\xspace}

\usepackage{cite} 
\usepackage{mciteplus}

\begin{document}

\renewcommand{\thefootnote}{\fnsymbol{footnote}}
\setcounter{footnote}{1}

\begin{titlepage}
\pagenumbering{roman}

\vspace*{-1.5cm}
\centerline{\large EUROPEAN ORGANIZATION FOR NUCLEAR RESEARCH (CERN)}
\vspace*{1.5cm}
\hspace*{-0.5cm}
\begin{tabular*}{\linewidth}{lc@{\extracolsep{\fill}}r}
\ifthenelse{\boolean{pdflatex}}
{\vspace*{-2.7cm}\mbox{\!\!\!\includegraphics[width=.14\textwidth]{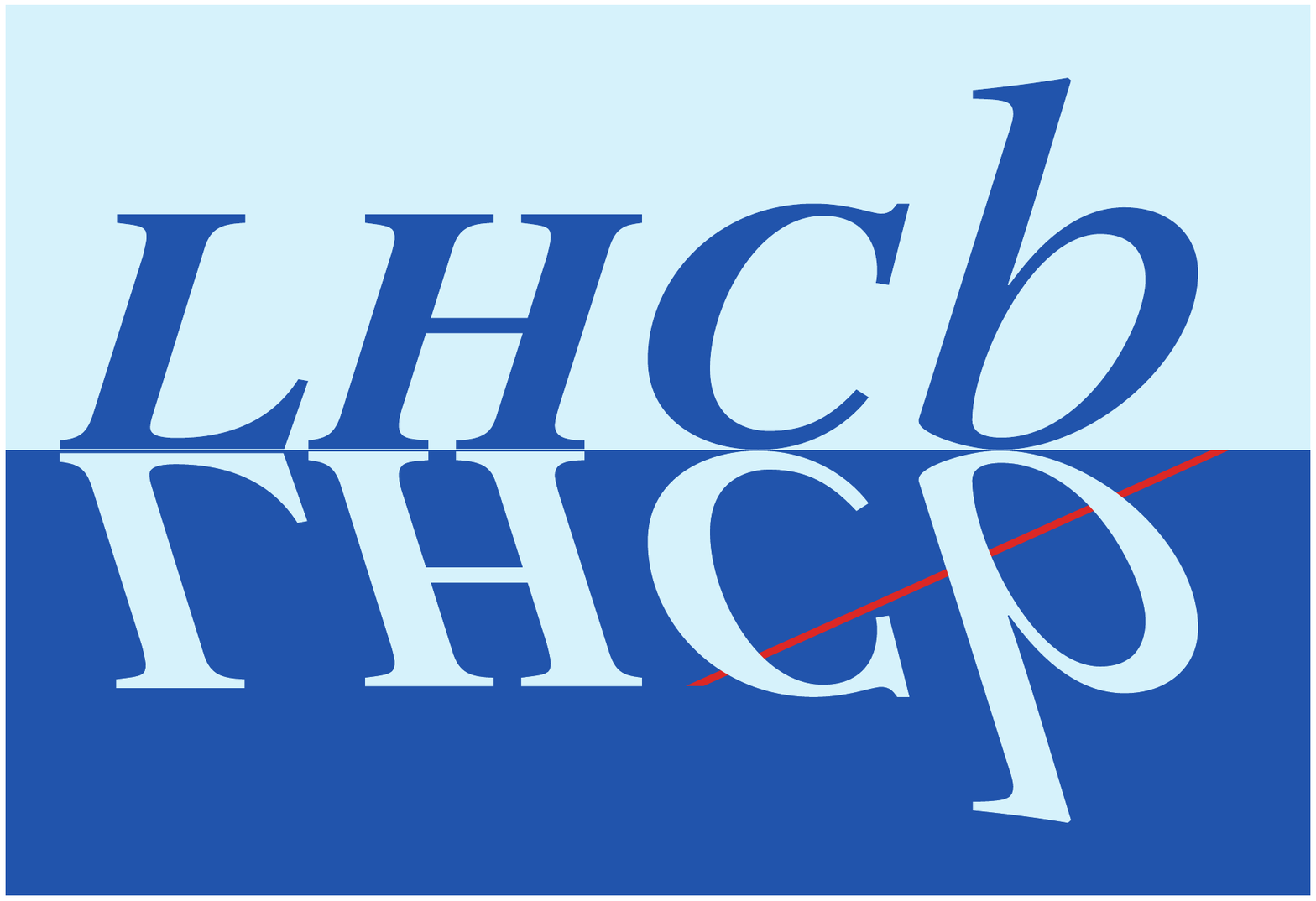}} & &}
{\vspace*{-1.2cm}\mbox{\!\!\!\includegraphics[width=.12\textwidth]{figs/lhcb-logo.eps}} & &}
\\
 & & CERN-PH-EP-2014-012 \\  
 & & LHCb-PAPER-2013-061 \\ 
 & & March 6, 2014 \\ 
 & & \\
\end{tabular*}

\vspace*{1.0cm}

{\bf\boldmath\Large
\begin{center}
  Searches for \Lb and \Xib decays to \KSppi and \KSpK final states with first observation of the \LbToKSppi decay
\end{center}
}

\vspace*{0.5cm}

\begin{center}
The LHCb collaboration\footnote{Authors are listed on the following pages.}
\end{center}

\vspace*{0.5cm}

\begin{abstract}
  \noindent
   A search for previously unobserved decays of beauty baryons to the final states \KSppi and \KSpK is reported.
   The analysis is based on a data sample corresponding to an integrated luminosity of $1.0 \invfb$ of $pp$ collisions.
   The \LbToKzbppi decay is observed with a significance of $8.6\,\sigma$, with branching fraction
\begin{eqnarray*}
  {\cal{B}}(\LbToKzbppi) & = & \left( 1.26 \pm 0.19 \pm 0.09 \pm 0.34 \pm 0.05 \right) \times 10^{-5} \,,
\end{eqnarray*}
where the uncertainties are statistical, systematic, from the ratio of fragmentation fractions $f_{\Lb}/f_{d}$, and from the branching fraction of the \BdToKzpipi normalisation channel, respectively.
A first measurement is made of the \CP asymmetry, giving
\begin{eqnarray*}
  A_{\CP}(\LbToKzbppi) & = & 0.22 \pm 0.13\stat \pm 0.03\syst \, .
\end{eqnarray*}
No significant signals are seen for \LbToKSpK decays, \Xib decays to both the \KSppi and \KSpK final states, and the \LbToDspToKsK decay, and upper limits on their branching fractions are reported. 
\end{abstract}

\vspace*{1.0cm}

\begin{center}
  Submitted to JHEP
\end{center}

\vspace{\fill}

{\footnotesize 
\centerline{\copyright~CERN on behalf of the \lhcb collaboration, license \href{http://creativecommons.org/licenses/by/3.0/}{CC-BY-3.0}.}}
\vspace*{2mm}

\end{titlepage}

\newpage
\setcounter{page}{2}
\mbox{~}
\newpage

\centerline{\large\bf LHCb collaboration}
\begin{flushleft}
\small
R.~Aaij$^{40}$, 
B.~Adeva$^{36}$, 
M.~Adinolfi$^{45}$, 
A.~Affolder$^{51}$, 
Z.~Ajaltouni$^{5}$, 
J.~Albrecht$^{9}$, 
F.~Alessio$^{37}$, 
M.~Alexander$^{50}$, 
S.~Ali$^{40}$, 
G.~Alkhazov$^{29}$
% ...

{\footnotesize \it
% ...
$ ^{5}$Clermont Universit\'{e}, Universit\'{e} Blaise Pascal, CNRS/IN2P3, LPC, Clermont-Ferrand, France\\
% ...
$ ^{9}$Fakult\"{a}t Physik, Technische Universit\"{a}t Dortmund, Dortmund, Germany\\
% ...
$ ^{29}$Petersburg Nuclear Physics Institute (PNPI), Gatchina, Russia\\
% ...
$ ^{36}$Universidad de Santiago de Compostela, Santiago de Compostela, Spain\\
$ ^{37}$European Organization for Nuclear Research (CERN), Geneva, Switzerland\\
% ...
$ ^{40}$Nikhef National Institute for Subatomic Physics, Amsterdam, The Netherlands\\
% ...
$ ^{45}$H.H. Wills Physics Laboratory, University of Bristol, Bristol, United Kingdom\\
% ...
$ ^{50}$School of Physics and Astronomy, University of Glasgow, Glasgow, United Kingdom\\
$ ^{51}$Oliver Lodge Laboratory, University of Liverpool, Liverpool, United Kingdom\\
% ...
}
\end{flushleft}

\cleardoublepage

\renewcommand{\thefootnote}{\arabic{footnote}}
\setcounter{footnote}{0}

\pagestyle{plain}
\setcounter{page}{1}
\pagenumbering{arabic}

\section{Introduction}
\label{sec:intro}

The study of beauty baryon decays is still at an early stage. 
Among the possible ground states with spin-parity $J^P = \frac{1}{2}^{+}$~\cite{PDG2012}, 
no hadronic three-body decay to a charmless final state has been observed. 
These channels provide interesting possibilities to study hadronic decays and to search for \CP violation effects, which may vary significantly across the phase-space~\cite{Bediaga:2009tr,Williams:2011cd}, as recently observed in charged \B meson decays to charmless three-body final states~\cite{LHCb-PAPER-2013-027,LHCb-PAPER-2013-051}.
In contrast to three-body neutral \B meson decays to charmless final states containing $\KS$ mesons~\cite{LHCb-PAPER-2013-042}, conservation of baryon number allows \CP violation searches without the need to identify the flavour of the initial state.

In this paper, a search is presented for \Lb and \Xib baryon decays to final states containing a \KS meson, a proton and either a kaon or a pion (denoted \LbXibToKSph where $h = \pi,K$).\footnote{
  The inclusion of charge-conjugate processes is implied throughout this paper, except where asymmetries are discussed. 
}
No published theoretical prediction or experimental limit exists for their branching fractions.
Intermediate states containing charmed hadrons are excluded from the signal sample and studied separately: the \LbToLcpiTopKS decay is used as a control channel, while the \LbToLcKTopKS and \LbToDspToKsK decays are also searched for.
The \LbToLcKToppiK decay has recently been observed~\cite{LHCb-PAPER-2013-056}, while the \LbToDsp decay has been suggested as a source of background to the $\Bs\to\Dsmp\Kpm$ mode~\cite{LHCb-PAPER-2011-022}.
All branching fractions are measured relative to that of the well-known control channel \BdToKzpipi~\cite{LHCb-PAPER-2013-042,Garmash:2006fh,Aubert:2009me}, relying on existing measurements of the ratio of fragmentation fractions $f_{\Lb}/f_d$, including its transverse momentum (\pt) dependence~\cite{LHCb-PAPER-2011-018,LHCb-PAPER-2012-037,LHCb-CONF-2013-011}.
When quoting absolute branching fractions, the results are expressed in terms of final states containing either \Kz or \Kzb mesons, according to the expectation for each decay, following the convention in the literature~\cite{PDG2012,HFAG}.

The paper is organised as follows.
A brief description of the LHCb detector and the data set used for the analysis is given in Sec.~\ref{sec:Det}.
The selection algorithms, the method to determine signal yields, and the systematic uncertainties on the results are discussed in Secs.~\ref{sec:sel}--\ref{sec:syst}.
The measured branching fractions are presented in Sec.~\ref{sec:results}.
Since a significant signal is observed for the \LbToKSppi channel, a measurement of its phase-space integrated \CP asymmetry is reported in Sec.~\ref{sec:results-acp}.
Conclusions are given in Sec.~\ref{sec:conclusions}.

\section{Detector and data set}
\label{sec:Det}

The \lhcb detector~\cite{Alves:2008zz} is a single-arm forward
spectrometer covering the \mbox{pseudorapidity} range $2<\eta <5$,
designed for the study of particles containing \bquark or \cquark
quarks. The detector includes a high precision tracking system
consisting of a silicon-strip vertex detector surrounding the $pp$
interaction region, a large-area silicon-strip detector located
upstream of a dipole magnet with a bending power of about
$4{\rm\,Tm}$, and three stations of silicon-strip detectors and straw
drift tubes placed downstream.
The combined tracking system provides momentum measurement with
relative uncertainty that varies from 0.4\% at 5\gevc to 0.6\% at 100\gevc, 
and impact parameter (IP) resolution of 20\mum for
tracks with high transverse momentum. 
Charged hadrons are identified using two ring-imaging Cherenkov (\rich) detectors~\cite{LHCb-DP-2012-003}. 
Photon, electron and hadron candidates are identified by a calorimeter system consisting of
scintillating-pad and preshower detectors, an electromagnetic
calorimeter and a hadronic calorimeter. Muons are identified by a
system composed of alternating layers of iron and multiwire
proportional chambers~\cite{LHCb-DP-2012-002}. 
The trigger~\cite{LHCb-DP-2012-004} consists of a
hardware stage, based on information from the calorimeter and muon
systems, followed by a software stage, which applies a full event
reconstruction.

The analysis is based on a sample, corresponding to an integrated luminosity
of $1.0\invfb$ of $pp$ collision data at a centre-of-mass energy of $7\tev$,
collected with the LHCb detector during 2011.
Samples of simulated events are also used to determine the signal
selection efficiency, to model signal event distributions and to
investigate possible background contributions.
In the simulation, $pp$ collisions are generated using
\pythia~6.4~\cite{Sjostrand:2006za} with a specific \lhcb
configuration~\cite{LHCb-PROC-2010-056}.  Decays of hadronic particles
are described by \evtgen~\cite{Lange:2001uf}, in which final-state
radiation is generated using \photos~\cite{Golonka:2005pn}. The
interaction of the generated particles with the detector and its
response are implemented using the \geant
toolkit~\cite{Allison:2006ve, *Agostinelli:2002hh} as described in
Ref.~\cite{LHCb-PROC-2011-006}.

\section{Selection requirements, efficiency modelling and background studies}
\label{sec:sel}

Events are triggered and subsequently selected in a similar way for both \LbXibToKSph signal modes and the \BdToKSpipi normalisation channel.
Events are required to be triggered at hardware level either by a calorimeter signal with transverse energy $\et >3.5 \gev$ associated with one of the particles in the signal decay chain, 
or by a particle in the event that is independent of the signal decay.
The software trigger requires a two-, three- or four-track
secondary vertex with a large sum of the transverse momentum of
the tracks and significant displacement from the primary $pp$
interaction vertices~(PVs). 
At least one track should have $\pt > 1.7\gevc$ and \chisqip with respect to any PV greater than 16, where \chisqip is defined as the difference in \chisq of a given PV reconstructed with and without the considered particle. 
A multivariate algorithm~\cite{BBDT} is used for the identification of secondary vertices consistent with the decay of a \bquark hadron.

An initial set of loose requirements is applied to filter the events selected by the trigger.     
Each \bquark hadron (\Lb, \Xib or \Bz) decay is reconstructed by combining two charged tracks with a \KS candidate.
The \KS candidates are reconstructed in the $\pip\pim$ final state, and are
classified into two categories. 
The first includes candidates that have hits in the vertex detector 
and the tracking stations downstream of the dipole magnet, hereafter referred to as ``\LL''.
The second category includes those decays in which track segments for the two pions are not found in the vertex detector, and use only the tracking stations downstream of the vertex detector (``\DD'').
The pions are required to have momentum $p > 2\gevc$ and to form a vertex with $\chisqvtx < 12$.
In addition, for \DD (\LL) \KS type the pions must have minimum \chisqip with respect to any PV greater than 4 (9), and the pair must satisfy $|m(\pip\pim) - m_{\KS}| < 30\,(20)\,\mevcc$, where $m_{\KS}$ is the known \KS mass~\cite{PDG2012}.
The \KS candidate is associated to the PV that minimises the \chisqip, and the square of the separation distance between the \KS vertex and the associated PV divided by its uncertainty (\chisqvs), must be greater than $50\,(90)$ for \DD (\LL) candidates.
For \DD \KS candidates $p > 6\,\gevc$ is also required.  

For both signal modes and the normalisation channel, the selection exploits the topology of the three-body decay and the $b$ hadron kinematic properties. 
The scalar sum of the transverse momenta of the daughters is required to be greater than $3\,\gevc$ and at least two of the daughters must have $\pt > 0.8\,\gevc$. 
The IP of the charged daughter with the largest \pt is required to be greater than $0.05 \mm$.
The minimum for each pair of two daughters of the square of the distance of closest approach divided by its uncertainty must be less than $5$. 
Furthermore, it is required that the $b$ hadron candidate has $\chisqvtx < 12$, $\chisqip < 4$, $\chisqvs > 50$, that its vertex separation from the PV must be greater than $1 \mm$, that the cosine of the ``pointing'' angle between its momentum vector and the line joining its production and decay vertices must be greater than 0.9999, and that it has $\pt > 1.5\,\gevc$.
Additional requirements are imposed to reduce background: the separation between the \KS and $b$ hadron candidate vertices must be positive in the $z$ direction;\footnote{The $z$ axis points along the beam line from the interaction region through the LHCb detector.} 
and the \KS flight distance must be greater than $15 \mm$.
The \bquark hadron candidates are required to have invariant mass within the ranges $5469 < m(\KS p h^{-}) < 5938 \mevcc$, evaluated for both $h=K,\pi$ hypotheses, and $4779 < m(\KS\pip\pim) < 5866 \mevcc$. 
To avoid potential biases during the selection optimisation, regions of $\pm 50\mevcc$ (\cf the typical resolution of 15\mevcc) around both the \Lb and \Xib known masses were not examined until the selection criteria were established.

Further separation of signal from combinatorial background candidates is achieved with a boosted decision tree (BDT) multivariate classifier~\cite{Breiman,AdaBoost}.
The BDT is trained using the \BdToKSpipi control channel as a proxy for the signal decays, with simulated samples used for the signal and data from the sideband region $5420 < m(\KS \pip\pim) < 5866 \mevcc$ for the background.
Potential baryonic contributions in the sidebands from \LbToKSppi and $\Lc\rightarrow \KS p$ decays are reduced by vetoing the relevant invariant masses in appropriate ranges.
In order to avoid bias in the training, the sample is split randomly into two, and two separate BDT trainings are used.
The set of input variables is chosen to optimise the performance of the algorithm, and to minimise efficiency variation across the phase-space. 
The input variables for the BDTs are the $\pt$, $\eta$, \chisqip, \chisqvs, pointing angle and \chisqvtx of the $b$ hadron candidate; the sum of the \chisqip values of the $h^{+}$ and $h^{-}$ tracks (here $h = \pi, K, \proton$); and the \chisqip, \chisqvs and \chisqvtx of the \KS candidate.

The choice of the optimal BDT cut value is determined separately for each \KS category, and separately for the charmless signal modes and for the channels containing intermediate \Lc or \Dsm hadrons.
An appropriate figure of merit for previously unobserved modes is~\cite{Punzi},
\begin{equation}
\label{eq:punzi}
{\cal{Q}} = \frac{\epsilon_{\rm sig}} {{a/2} + \sqrt{B}}\,,
\end{equation}
where $a=5$ quantifies the target level of significance in units of standard deviations, $\epsilon_{\rm sig}$ is the efficiency of the signal selection determined from the simulation, and $B$ is the expected number of background events in the signal region, which is estimated by extrapolating the result of a fit to the invariant mass distribution of the data sidebands. 
An alternative optimisation approach, which minimises the expected upper limit~\cite{LHCb-PAPER-2013-029}, is also investigated and provides a similar result. 

Potential sources of remaining background are suppressed with particle identification (PID) criteria.
This is of particular importance for reducing cross feed between the signal channels due to kaon/pion misidentification.
Particle identification information is provided by the \rich detectors~\cite{LHCb-DP-2012-003}, in terms of the logarithm of the likelihood ratio between the kaon/proton and pion hypotheses (DLL$_{K\pi}$ and DLL$_{p\pi}$). 
A tight DLL$_{p\pi}$ criterion on the proton candidate suppresses most possible backgrounds from misidentified \bquark hadron decays. 
An additional DLL$_{K\pi}$ requirement is imposed to reduce cross feed between $\KS p\pim$ and $\KS p\Km$ modes.
In addition, candidates containing tracks with associated hits in the muon detectors are rejected.
The DLL requirements are optimised using Eq.~(\ref{eq:punzi}), and their efficiencies are determined using high-purity data control samples of $\Lz \rightarrow p\pi^{-}$ and $\Dz \rightarrow \Km\pip$ decays, reweighted according to the expected signal kinematic (momentum and \pt) distributions from the simulation.

The efficiency of the selection requirements is studied with simulation.
A multibody decay can in general proceed through intermediate states and through a nonresonant amplitude.  
It is therefore necessary to model the variation of the efficiency, and to account for the distribution of signal events, over the phase-space of the decay.
The phase-space of the decay of a spin-zero particle to three spin-zero particles can be completely described by the Dalitz plot~\cite{Dalitz:1953cp} of any pair of the two-body invariant masses squared.
The situation for a baryon decay is more complicated due to the spins of the initial and final state fermions, but the conventional Dalitz plot can still be used if spin effects are neglected.\footnote{
  Note that \Lb baryons produced in $pp$ collisions at $\sqrt{s} = 7\tev$ have been measured to have only a small degree of polarisation~\cite{LHCb-PAPER-2012-057}.
}
For three-body \bquark hadron decays, both signal decays and the dominant combinatorial backgrounds populate regions close to the kinematic boundaries of the conventional Dalitz plot.  
For more accurate modelling of those regions, it is convenient to transform to a rectangular space (hereafter referred to as the square Dalitz plot~\cite{Aubert:2005sk}) described by the variables $m^\prime$ and $\theta^\prime$ where 
\begin{equation}
\label{eq:sqdp-vars}
m^\prime \equiv \frac{1}{\pi}
\arccos\left(2\frac{m(\KS\proton) - m^{\rm min}(\KS\proton)}{m^{\rm max}(\KS\proton) - m^{\rm min}(\KS\proton)} - 1 \right)\,,
\hspace{7mm}
\theta^\prime \equiv \frac{1}{\pi}\theta(\KS\proton)\,.
\end{equation}
Here $m(\KS\proton)$ is the invariant mass of the \KS and proton, $m^{\rm max}(\KS\proton) = m_{\Lb} - m_{h^-}$ and $m^{\rm min}(\KS\proton) = m_{\KS} + m_{\proton}$ are the boundaries of $m(\KS\proton)$, $\theta(\KS\proton)$ is the angle between the $\proton$ and the $h^-$ track in the $\KS\proton$ rest frame.

Simulated events are binned in the square Dalitz plot variables in order to determine the selection efficiencies.
If no significant $b$ hadron signal is seen, the efficiency corresponding to a uniform distribution across the square Dalitz plot is used as the nominal value, and a systematic uncertainty is assigned due to the variation across the phase-space.
When the signal yield has significance (evaluated as described in the next section) greater than $3\,\sigma$, the signal distribution in the square Dalitz plot is obtained with the \sPlot\ technique~\cite{Pivk:2004ty} (with the \bquark hadron candidate invariant mass used as the control variable), and the efficiency corresponding to the observed distribution is used.

There is limited prior knowledge of the branching fractions of $b$ baryon decays that may form backgrounds to the current search. 
Numerous modes are investigated with simulation, and the only significant potential background contribution that is found to peak in the candidate mass distribution is from \LbToLchToppiK decays, where the kaon is misidentified as a pion, and the $\pi K$ pair can form a \KS candidate. 
To suppress this background, candidates that have $\proton \Km \pip$ masses within $30\mevcc$ of the known \Lc mass are vetoed.

The decays \LbToLchTopKS and \LbToDspToKsK share the same final state as the charmless signal modes and are removed by vetoing regions in $m(\KS\proton)$ and $m(\KS K)$ within $\pm 30\mevcc$  of the known $\Lc$ and $D^{-}_{s}$ masses.
These vetoes are reversed to select and study the decay modes with intermediate charmed states.
The additional requirement for the charmed modes reduces the combinatorial background.
Therefore the optimal BDT requirement is obtained separately for each channel.

The backgrounds to the normalisation channel are treated as in Ref.~\cite{LHCb-PAPER-2013-042}.
The main contributions are considered to be charmless decays with an unreconstructed photon in the final state (\eg\ \BdToKsPiPiGamma or $\Bd\to\etapr(\to\rho^0\gamma)\KS$), charmless decays of \Bd or \Bu mesons into two vector particles (\eg\ \BdToKstrho and \BuToKstrho) where a soft pion is not reconstructed, and charmed decays (\eg\ \BuToDzpi) where a pion is not reconstructed.

\section{Fit model and results}
\label{sec:fit}

All signal and background yields are determined simultaneously by performing an unbinned extended maximum likelihood fit to the \bquark hadron candidate invariant mass distribution of each final state and \KS category.
The probability density function (PDF) in each invariant mass distribution is defined as the sum of several components (signal, cross-feed contributions, combinatorial and other backgrounds), with shapes derived from simulation.

Signal PDFs are known to have asymmetric tails that result from a combination of the effects of final state radiation and stochastic tracking imperfections.
The \LbXibToKSph signal mass distributions
are modelled by the sum of a ``core'' Gaussian and a bifurcated Gaussian function, that share the same mean value.
The core resolution is allowed to be different for each \KS category, whilst the two widths of the bifurcated Gaussian are common to \DD and \LL types.
Alternative shapes are studied using simulation, and this choice is found to provide the most stable and accurate description for a given number of parameters.

The significant yield of \LbToLcpiTopKS decays allows a subset of fit parameters common to the unobserved $b$ baryon decays to be determined from data. 
The core width and the relative fraction between the Gaussian and bifurcated Gaussian component are therefore expressed in terms of the parameters obtained from the fit to \LbToLcpiTopKS candidates, with deviations from those values allowed within ranges as seen in the simulation.
Explicitly, the function used for each unobserved channel $j$ and \KS type $c$ is 
\begin{equation}
\label{eq:pdf_signals}
{\rm PDF}(m;\mu,\sigma^{c}_{\rm core},\sigma_{\rm R},\sigma_{\rm L}) = s^{c,j}_{f}f^{c} G(m;\mu,s^{c,j}_{\sigma}\sigma^{c}_{\rm core}) + (1 - s^{c,j}_{f}f^{c} ) 
B(m;\mu,\sigma_{\rm L},\sigma_{\rm R})
, 
\end{equation}
where $m$ is the invariant mass of the \bquark hadron candidate and
$G$ and $B$ represent the Gaussian and bifurcated Gaussian distributions respectively.
The parameters $\sigma_{\rm L}$ and $\sigma_{\rm R}$ are respectively the left and right widths of the bifurcated Gaussian function, 
$\sigma^{c}_{\rm core}$ and $f^{c}$ are the width and the fraction of the core Gaussian for \LbToLcpiTopKS candidates,
while $s^{c,j}_{\sigma}$ and $s^{c,j}_{f}$ are the corresponding scale factors for the channel $j$, determined from simulation.
The peak position $\mu$ for \Lb decays is shared among all modes, while that for \Xib decays is fixed according to the measured \Lb and \Xib mass difference, $m_{\Xib} - m_{\Lb} = 168.6 \pm 5.0 \mevcc$~\cite{PDG2012}.
The scale factors for \Lb and \Xib signal shapes are allowed to differ but are found to be consistent.   
The fit model and its stability are validated with ensembles of pseudo-experiments, and no significant bias is found.

The normalisation channel is parametrised following Ref.~\cite{LHCb-PAPER-2013-042}.
The signal distribution of the \B candidate invariant mass is modelled by the sum of two Crystal Ball (CB) functions~\cite{Skwarnicki:1986xj}, 
where the power law tails are on opposite sides of the peak. 
The two CB functions are constrained to have the same peak position and resolution, which are floated in the fit.
The tail parameters and the relative normalisation of the two CB functions are taken from the simulation and fixed in the fit to data.
To account for \BsToKSpipi decays~\cite{LHCb-PAPER-2013-042} an additional component, parametrised in the same way as the \Bz channel, is included.
Its peak position is fixed according to the known $\Bs-\Bd$ mass difference~\cite{PDG2012}, its width is constrained to be the same as that seen for the \Bd mode to within the difference found in simulation, and its yield is allowed to vary independently.

An exponential shape is used to describe the combinatorial background, which is treated as independent for each decay mode and \KS type.
Cross-feed contributions are also considered for each $\KS ph^{-}$ final state.
For the normalisation channel, a contribution from \BsToKSKpi decays is included, while yields of other possible misidentified backgrounds are found to be negligible~\cite{LHCb-PAPER-2013-042}.
Cross-feed and misidentified \BsToKSKpi shapes are modelled by double CB functions, with independent peak positions and resolutions.
The yields of these components are constrained to be consistent with the number of signal candidates in the corresponding correctly identified spectrum, multiplied by the relevant misidentification probability.
The peaking backgrounds to the normalisation channel reported in Sec.~\ref{sec:sel} are modelled by a generalised ARGUS function~\cite{Albrecht:1990cs} 
convolved with a Gaussian function with width determined from simulation. 
The yield of each contribution is constrained within uncertainty according to the corresponding efficiency and branching fraction.

\begin{figure}[htb]
\centering
\includegraphics*[width=0.46\textwidth]{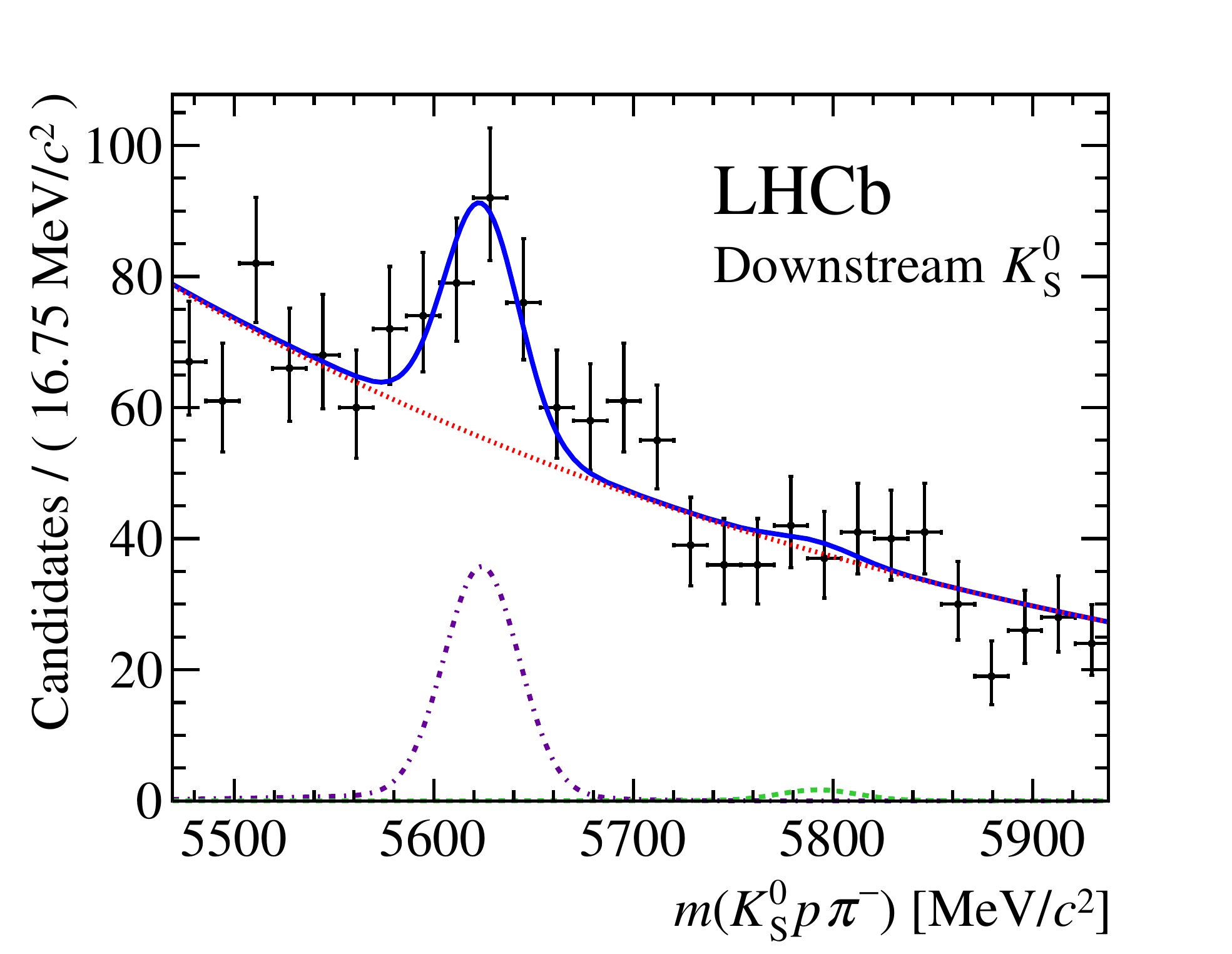}
\includegraphics*[width=0.46\textwidth]{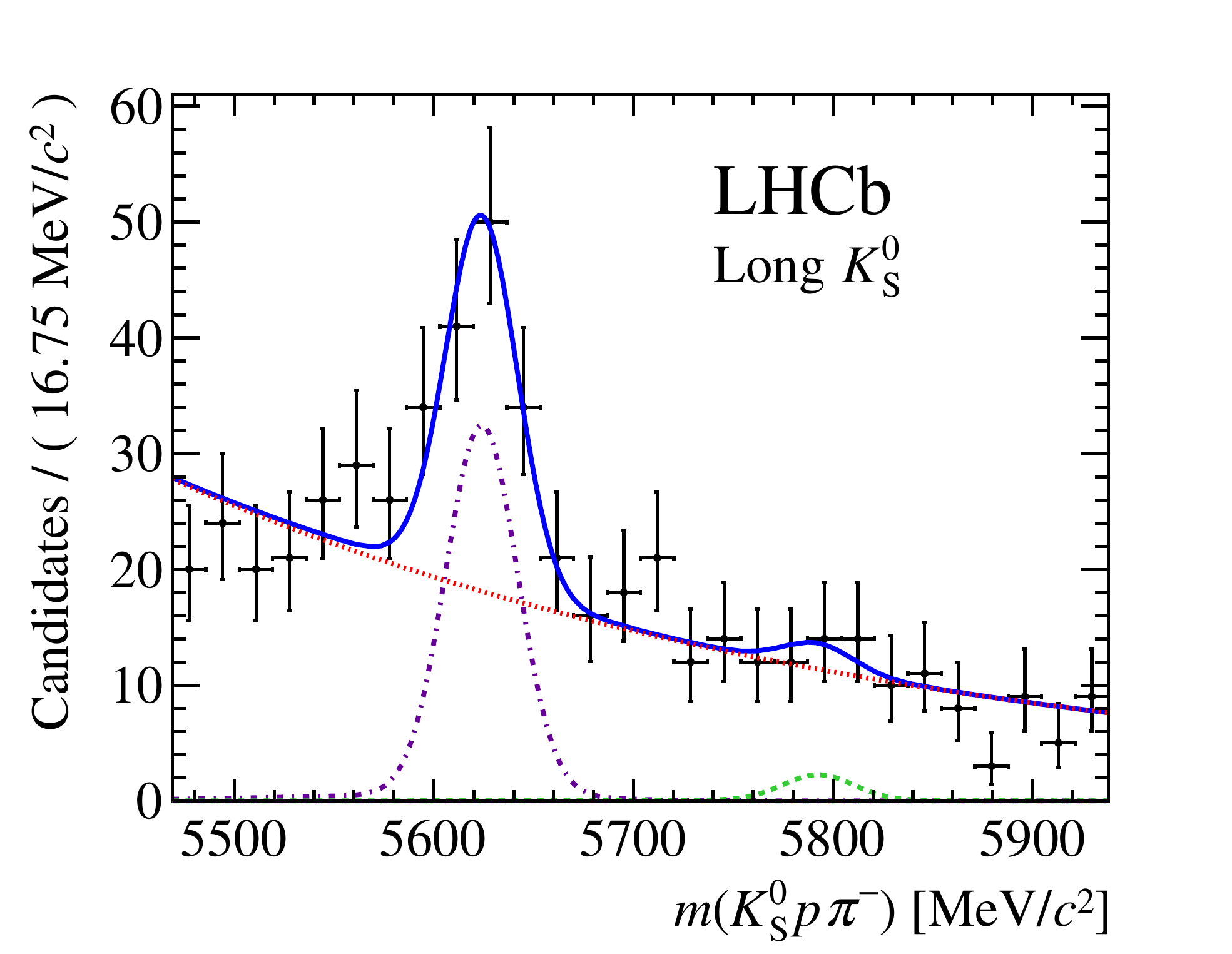}
\includegraphics*[width=0.46\textwidth]{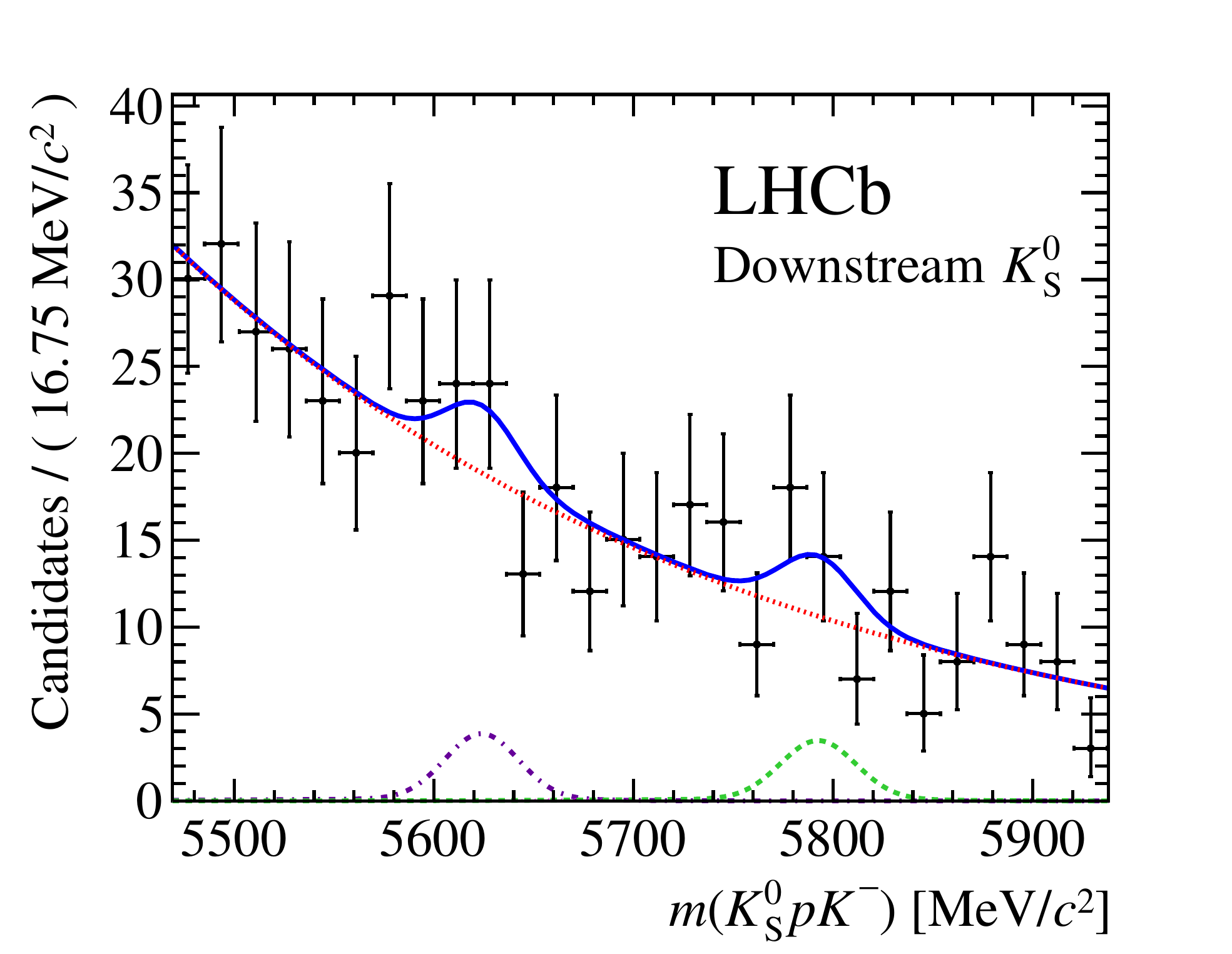}
\includegraphics*[width=0.46\textwidth]{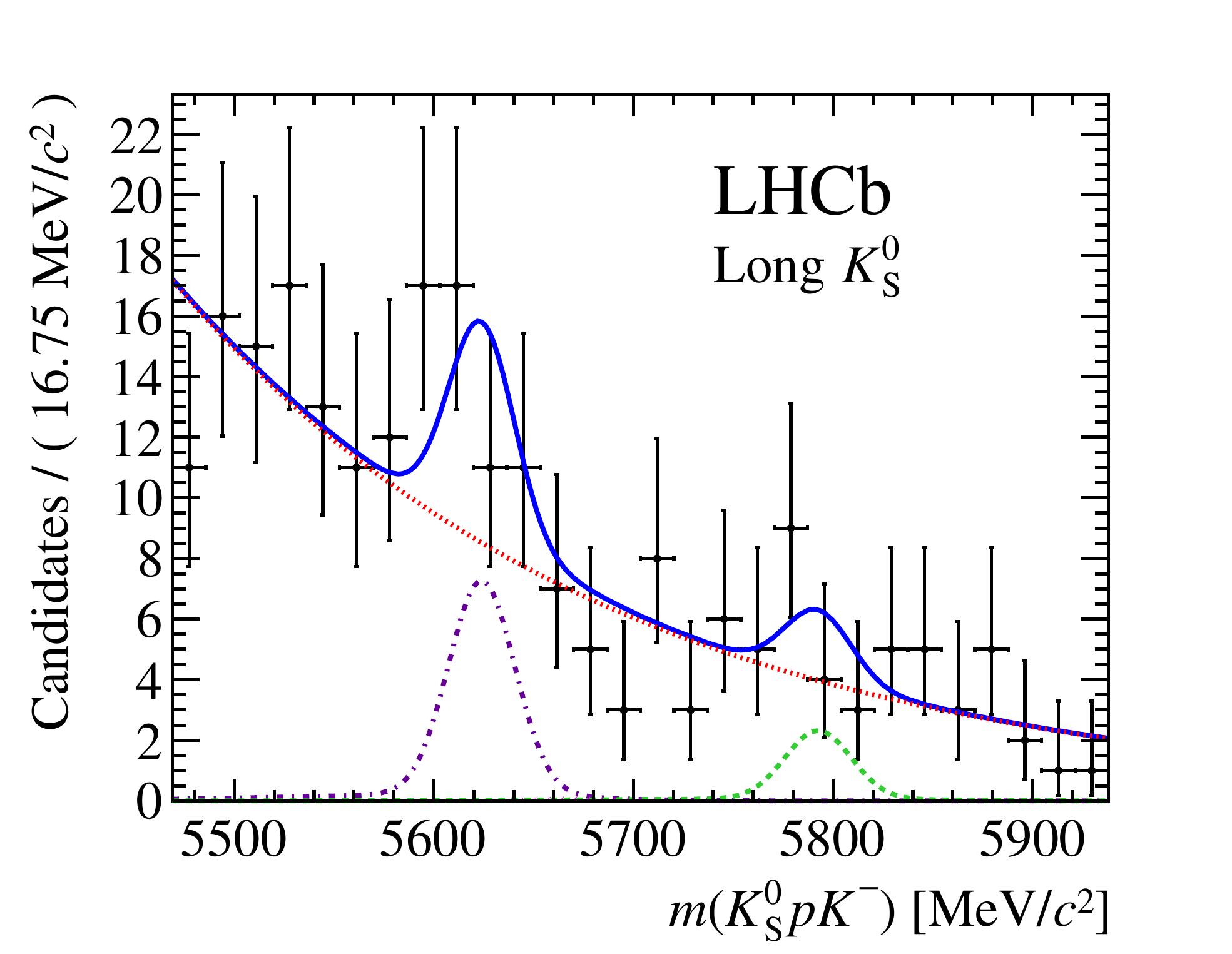}
\caption{\small 
  Invariant mass distribution of (top) $\KS p\pi^{-}$ and (bottom) $\KS pK^{-}$ candidates for the (left) \DD and (right) \LL \KS categories after the final selection in the full data sample.
  Each significant component of the fit model is displayed: \Lb signal (violet dot-dashed), $\Xib$ signal (green dashed)
  and combinatorial background (red dotted).  The overall fit is given by the solid blue line.
  Contributions with very small yields are not shown.
}
\label{fig:fit_KSppi}
\end{figure}

\begin{figure}[htb!]
\centering
\includegraphics*[width=0.46\textwidth]{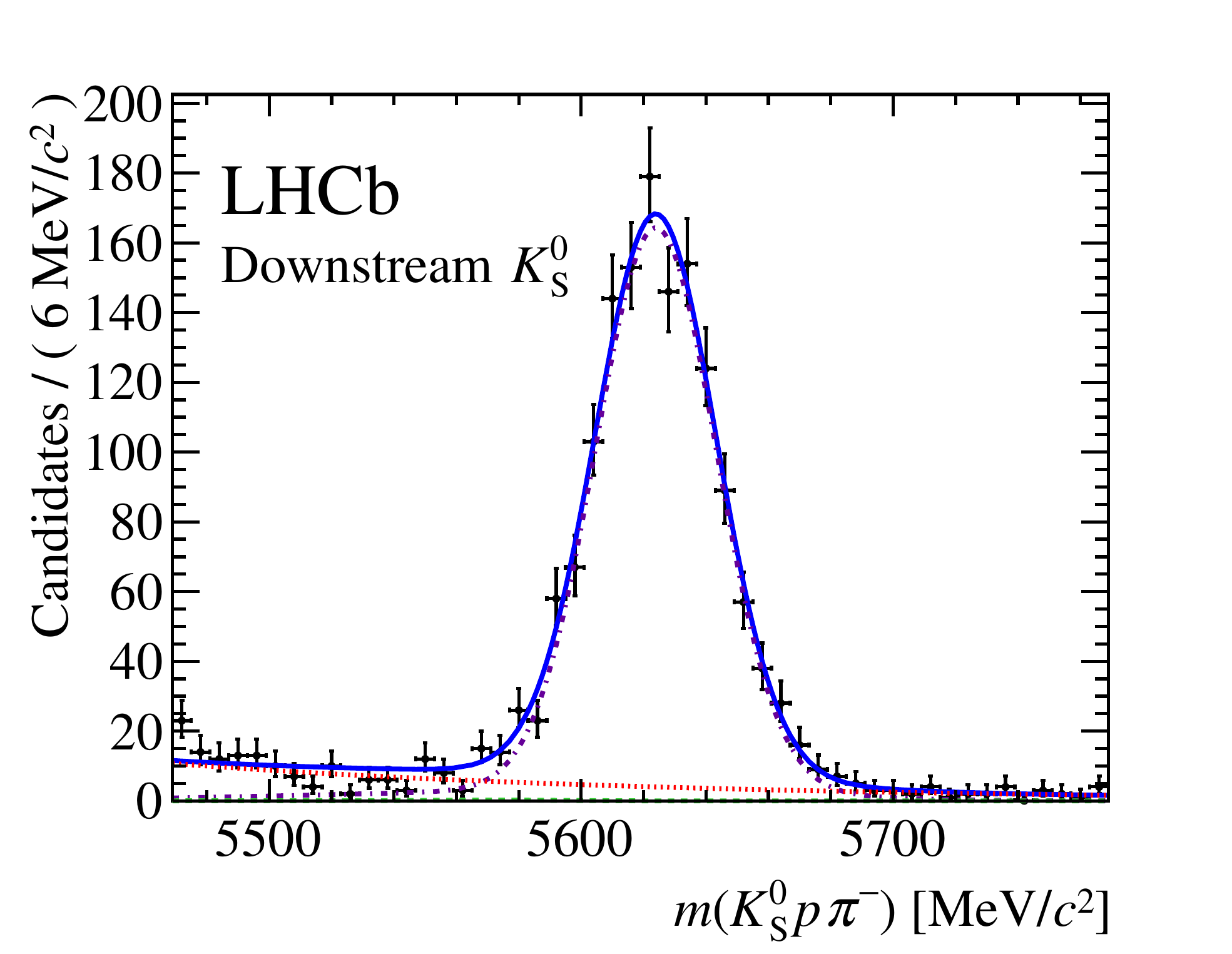}
\includegraphics*[width=0.46\textwidth]{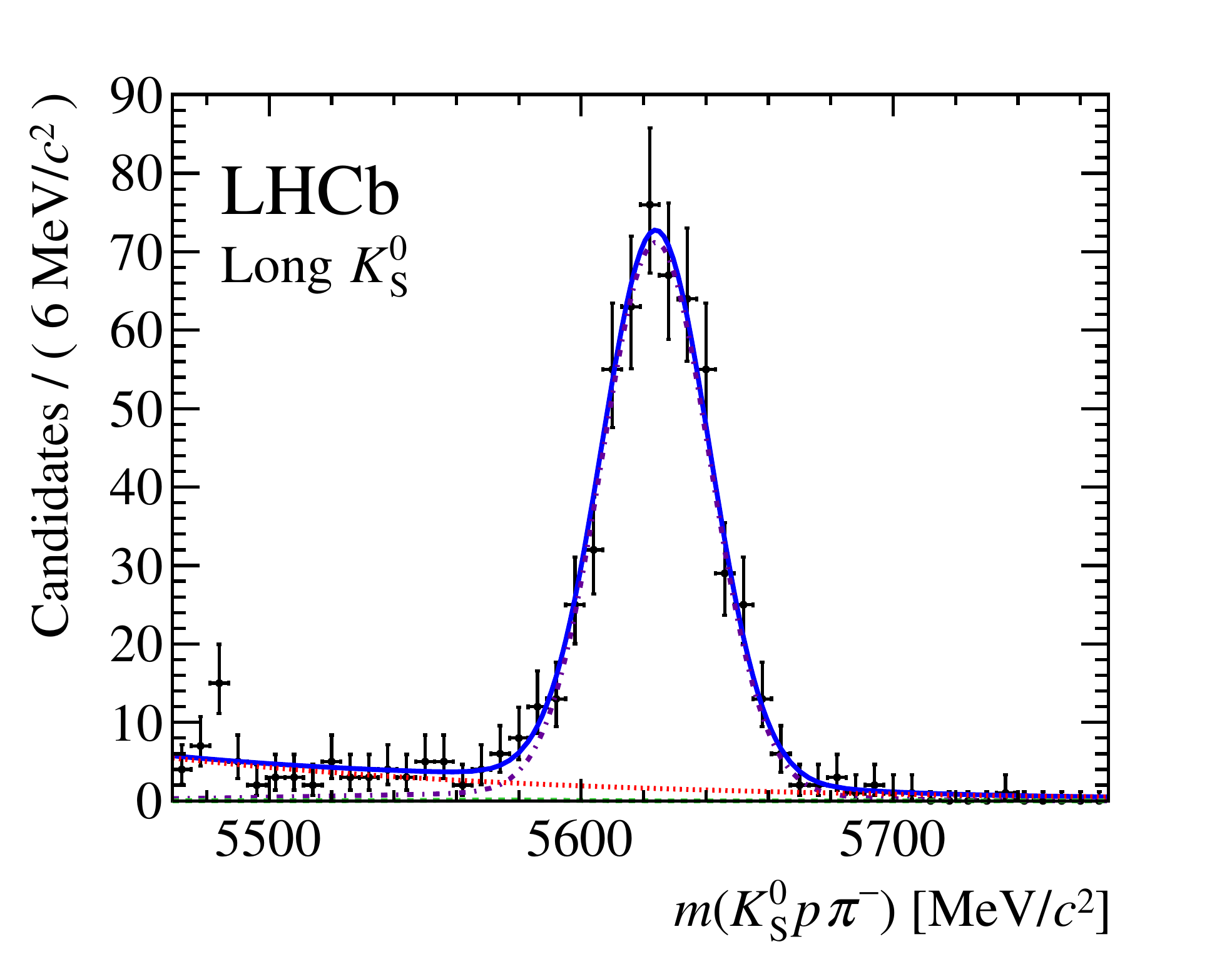}
\includegraphics*[width=0.46\textwidth]{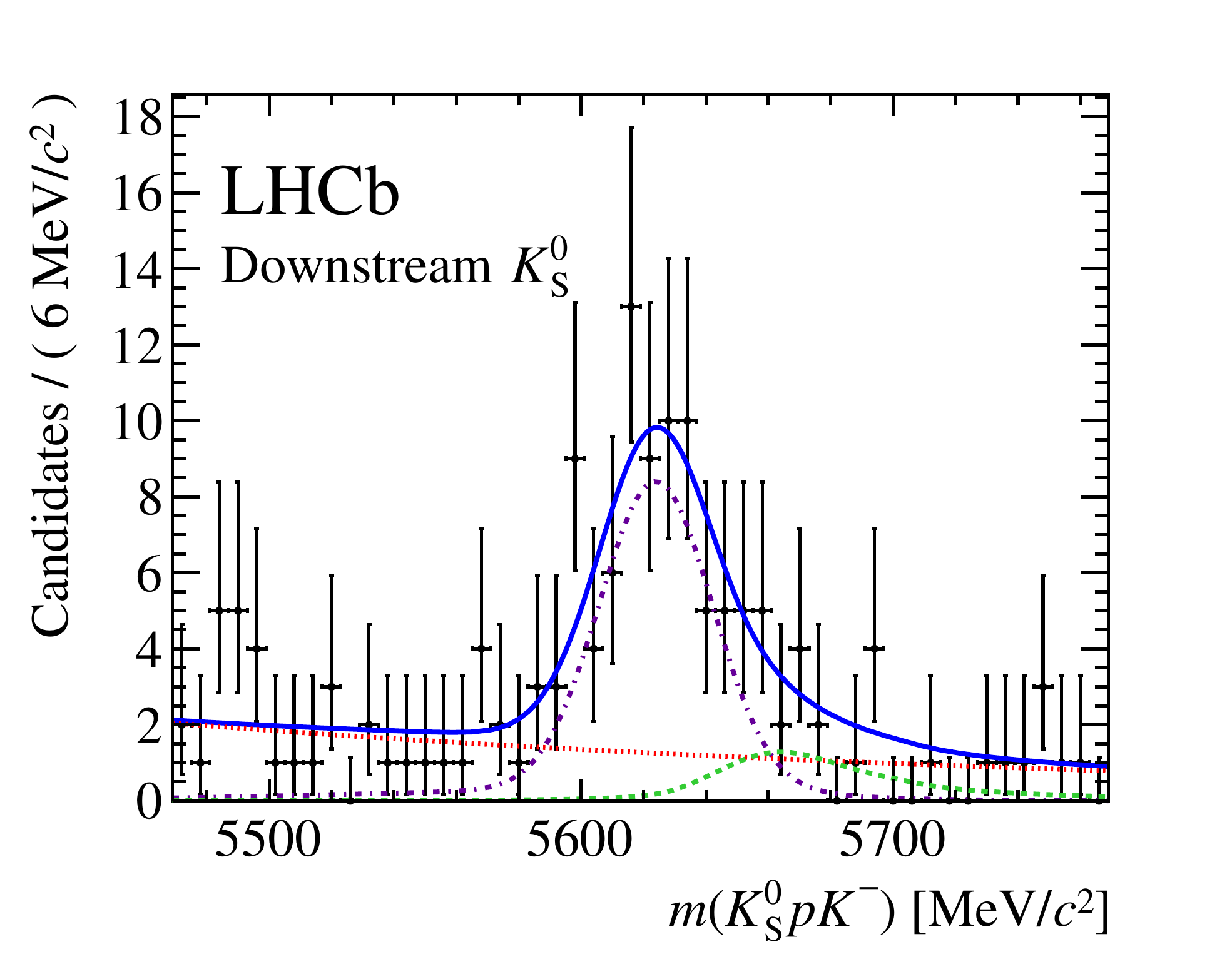}
\includegraphics*[width=0.46\textwidth]{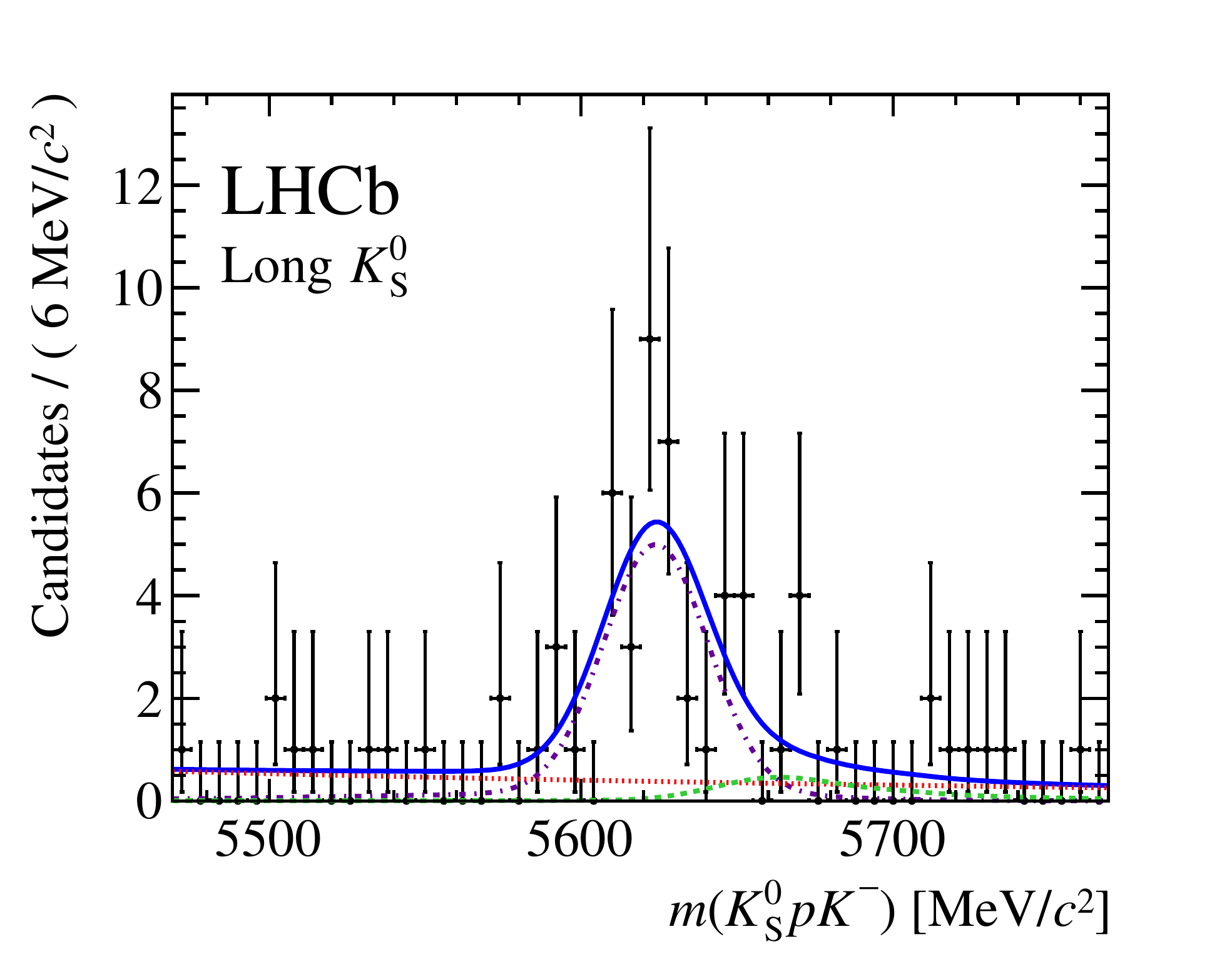}
\includegraphics*[width=0.46\textwidth]{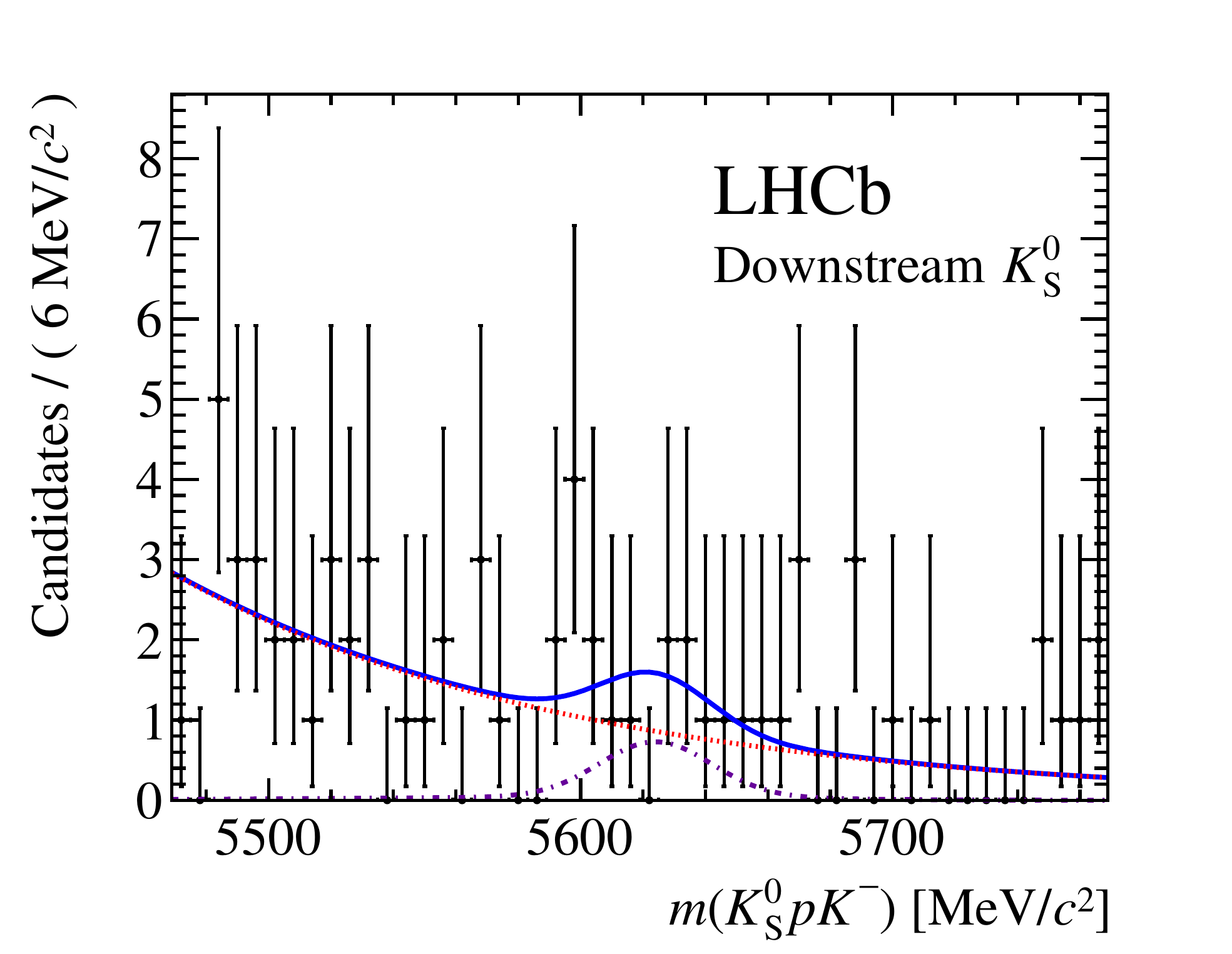}
\includegraphics*[width=0.46\textwidth]{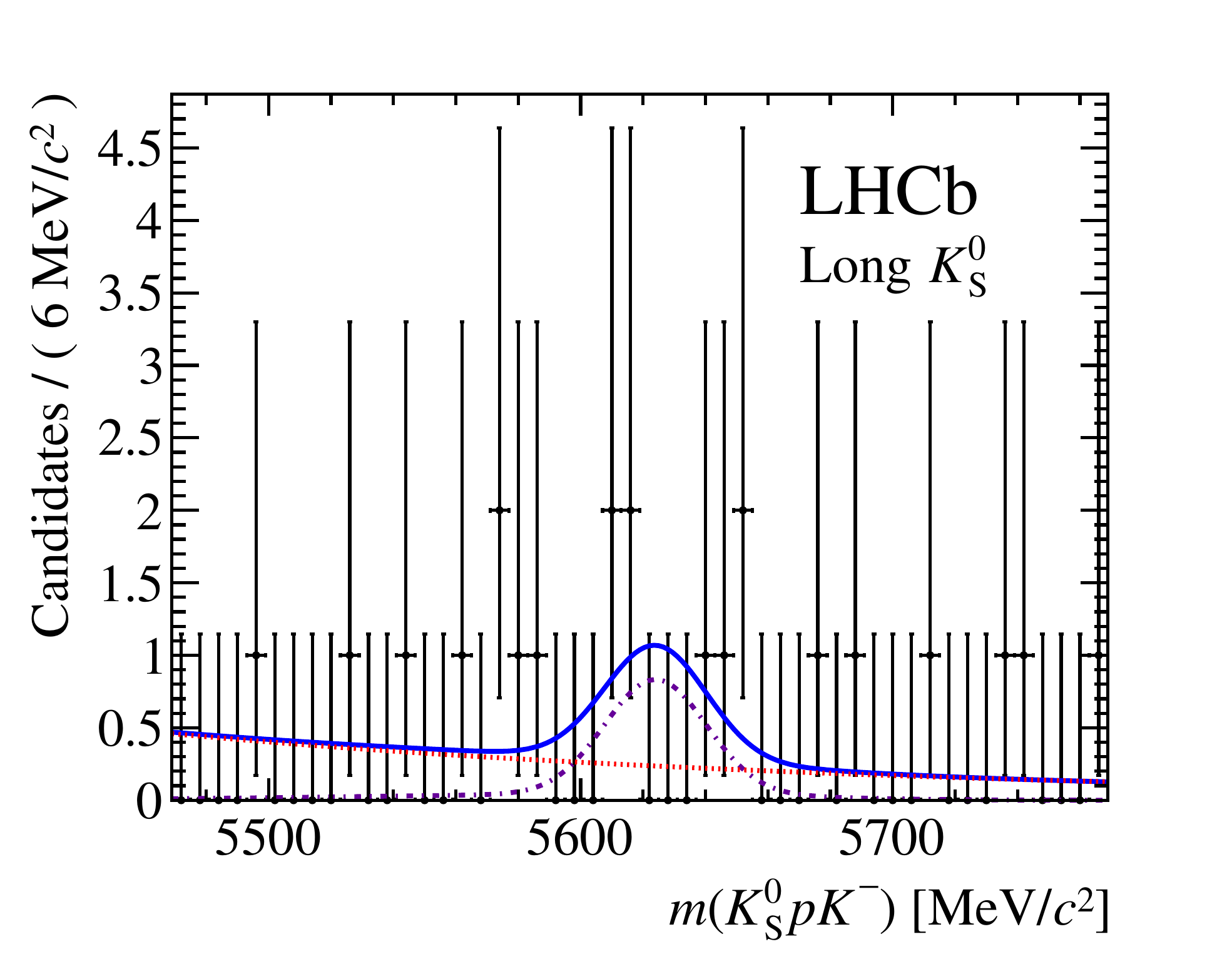}
\caption{\small 
  Invariant mass distribution of (top) \LbToLcpiTopKS, (middle) \LbToLcKTopKS and (bottom) \LbToDspToKsK candidates for the (left) \DD and (right) \LL \KS categories after the final selection in the full data sample.
  Each significant component of the fit model is displayed: signal PDFs (violet dot-dashed), signal cross-feed contributions (green dashed) and combinatorial background (red dotted). 
  The overall fit is given by the solid blue line.
  Contributions with very small yields are not shown.
}
\label{fig:fit_charmed}
\end{figure}

\begin{figure}[htb]
\centering
\includegraphics*[width=0.45\textwidth]{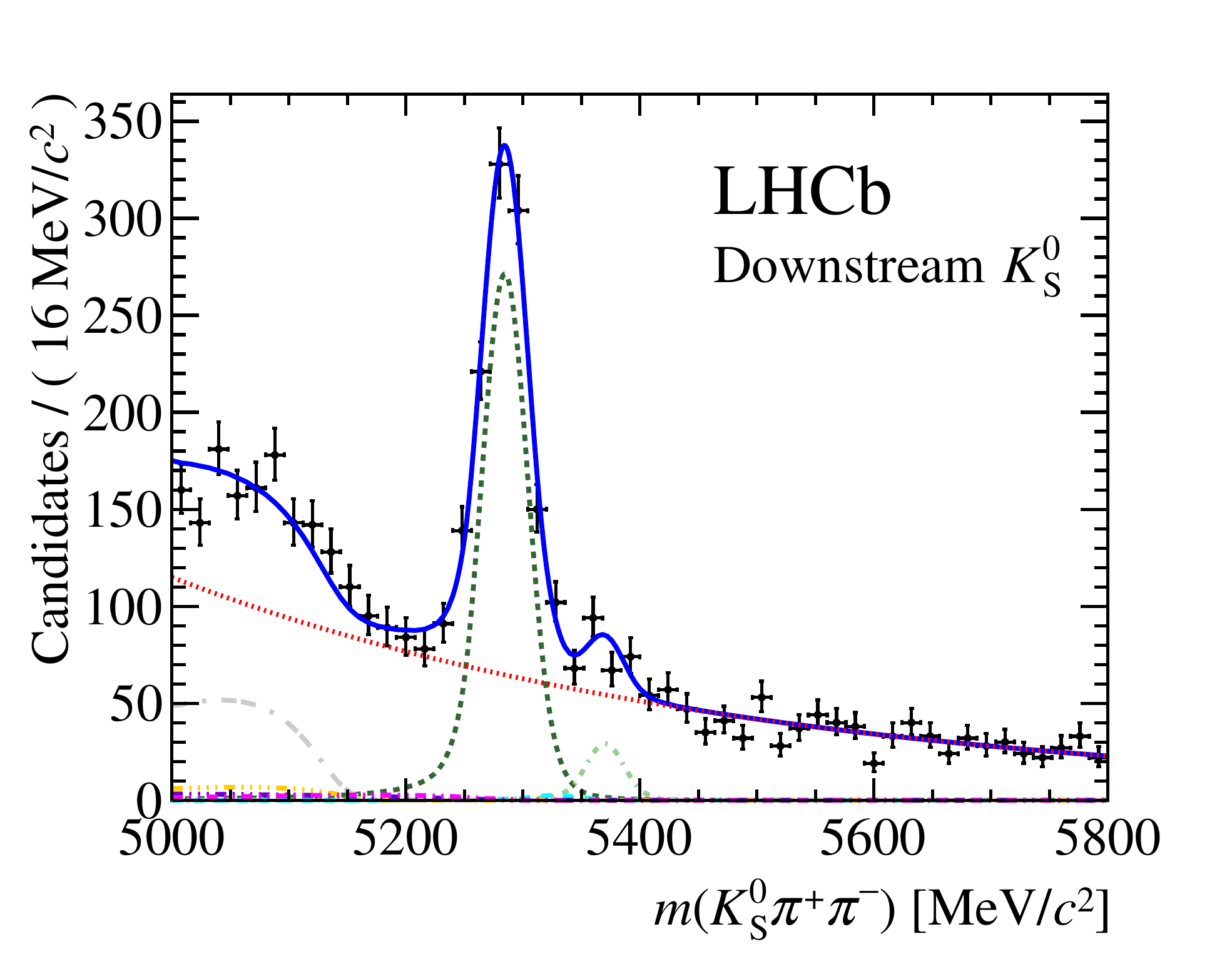}
\includegraphics*[width=0.45\textwidth]{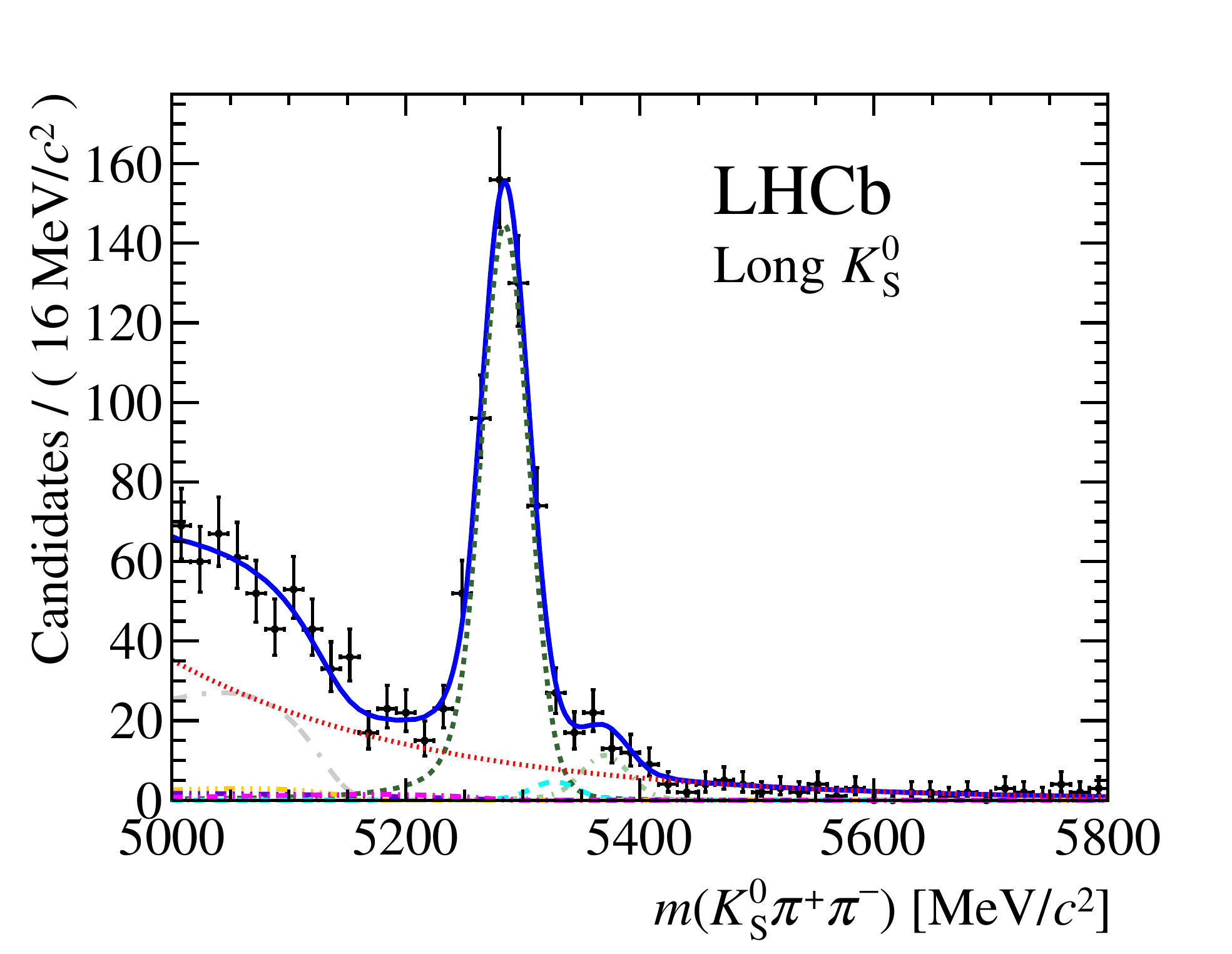}
\includegraphics*[width=0.45\textwidth]{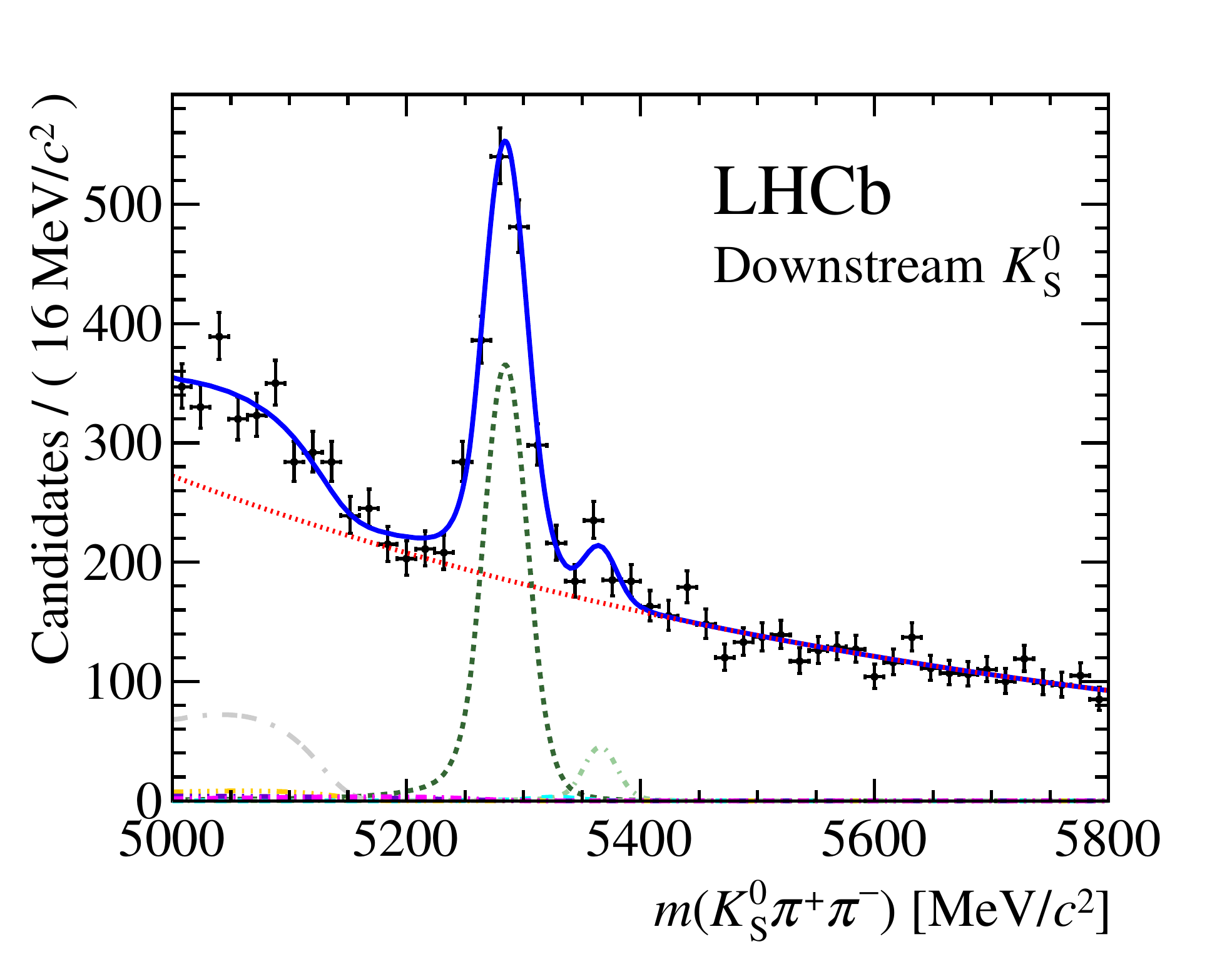}
\includegraphics*[width=0.45\textwidth]{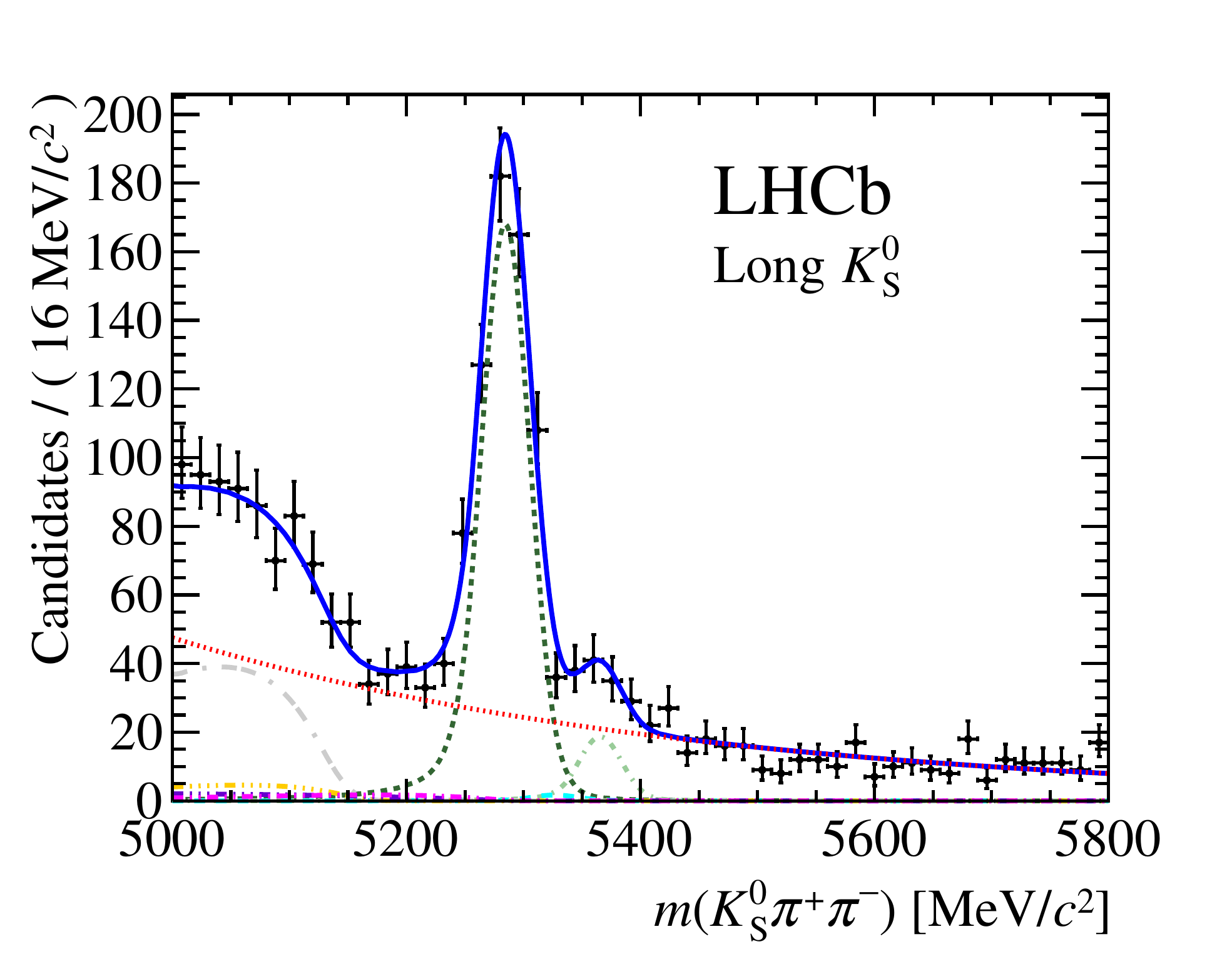}
\includegraphics*[width=0.45\textwidth]{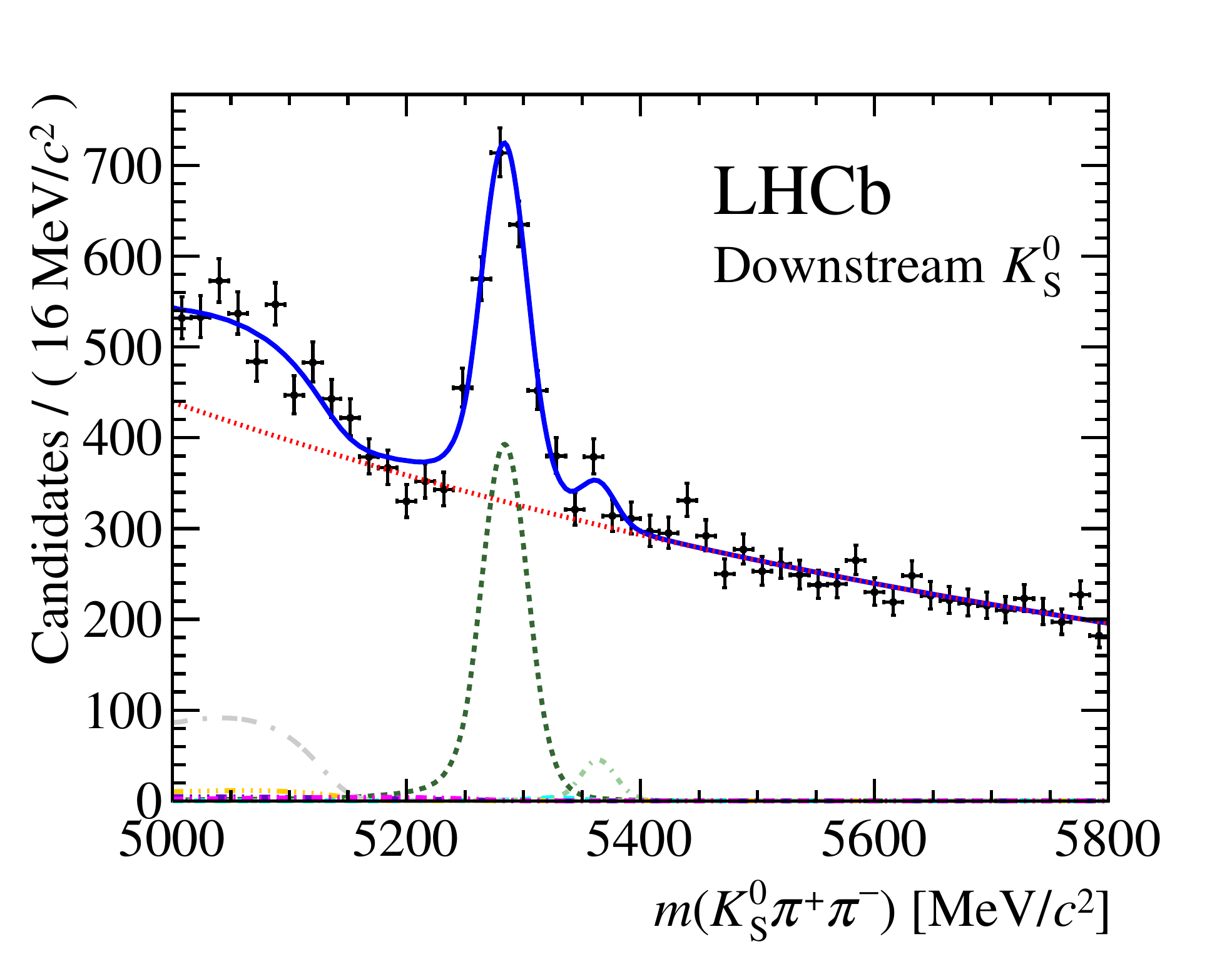}
\includegraphics*[width=0.45\textwidth]{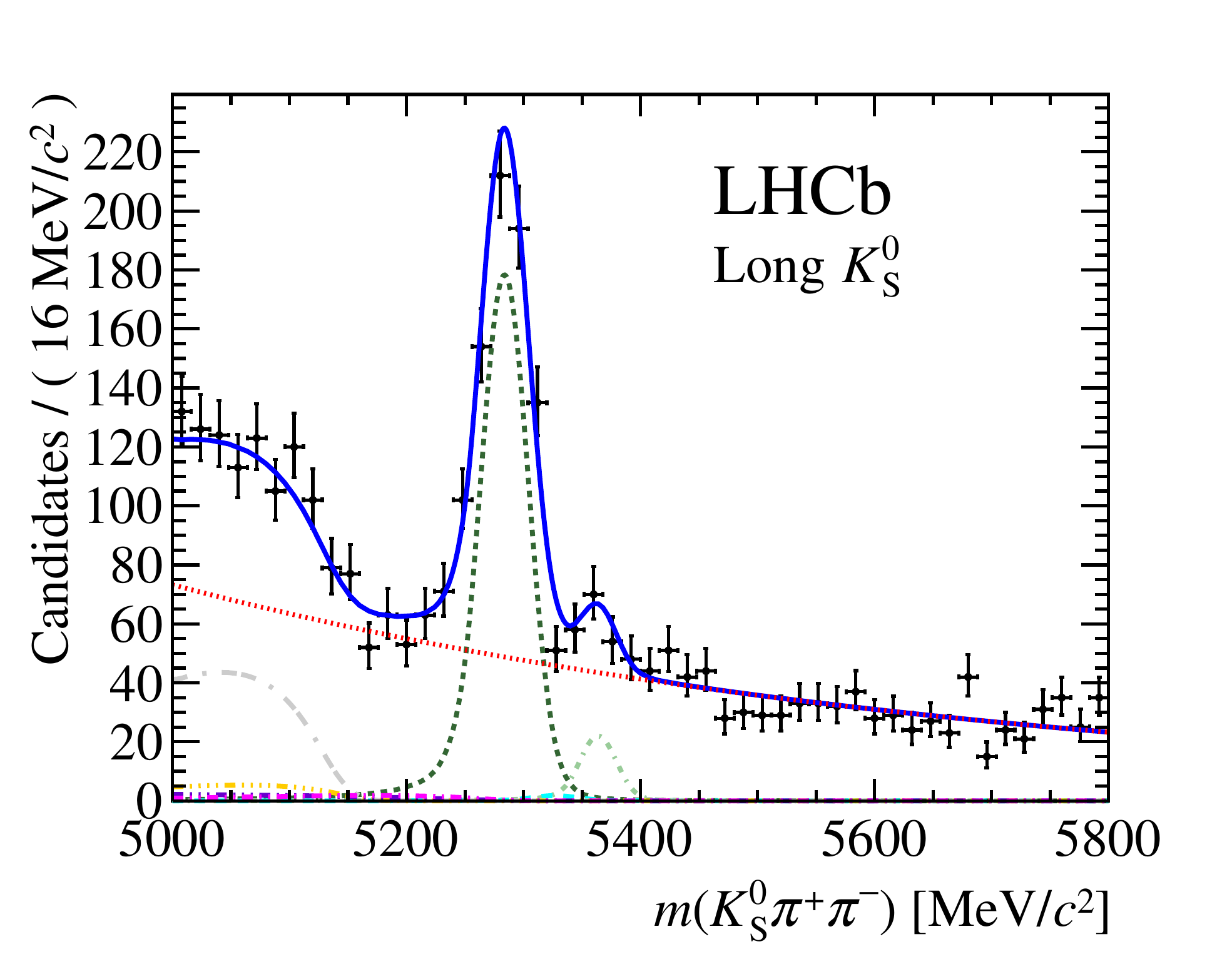}
\caption{\small 
  Invariant mass distribution of $\KS \pi^{+}\pi^{-}$ candidates with the selection requirements for the (top) \LbToKSph, (middle) \LbToLchTopKS and (bottom) \LbToDsp channels separated into (left) \DD and (right) \LL \KS categories. 
  Each component of the fit model is displayed: the \Bd (\Bs) decay is represented by the dashed dark (dot dashed light) green line; the background from \BsToKSKpi decays by the long dashed cyan line; \BuToDzpi (grey double-dash dotted), charmless \Bz(\Bp) decays (orange dash quadruple-dotted), \BdToetapKs (magenta dash double-dotted) and \BdToKsPiPiGamma (dark violet dash triple-dotted) backgrounds; the overall fit is given by the solid blue line; and the combinatorial background by the dotted red line.
}
\label{fig:fit_KSpipi}
\end{figure}

The results of the fit to data are shown in Fig.~\ref{fig:fit_KSppi} for
\LbXibToKSph candidates, Fig.~\ref{fig:fit_charmed} for \LbToLchTopKS and
\LbToDsp candidates and Fig.~\ref{fig:fit_KSpipi} for the \BdToKSpipi
normalisation channel, separated by \KS type. 
The fitted yields and relevant efficiencies are gathered in Table~\ref{tab:fit_results}.
The statistical significance of each signal is computed as $\sqrt{ 2 \ln (L_{\rm sig}/L_{0} )}$, where $L_{\rm sig}$ and $L_{0}$ are the likelihoods from the nominal fit and from the fit omitting the signal component, respectively.
These statistical likelihood curves for each \KS category are convolved with a Gaussian function of width given by the systematic uncertainty on the fit yield.
The total significance, for \DD and \LL \KS types combined, is found to be $8.6\,\sigma$ and $2.1\,\sigma$ for \LbToKSppi and \LbToKSpK decays, respectively.
Moreover, the statistical significance for the \LbToLcKTopKS decay is found to be $9.4\,\sigma$ and $8.0\,\sigma$ for \DD and \LL categories respectively, confirming the recent observation of this channel~\cite{LHCb-PAPER-2013-056}. 
The significances of all other channels are below $2\,\sigma$.

\begin{figure}[htb]
\centering
\includegraphics*[width=0.6\textwidth]{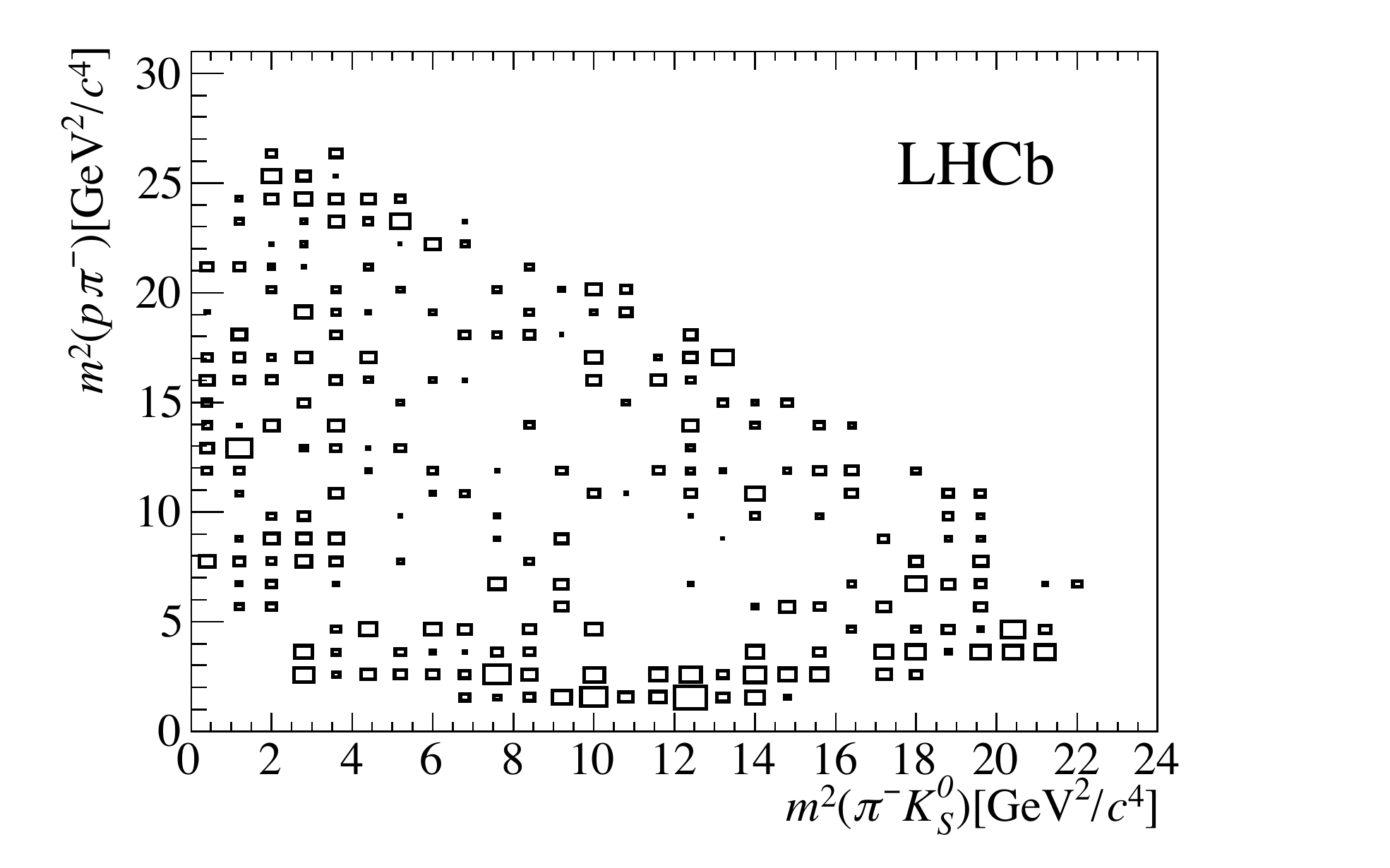}
\caption{\small 
  Background-subtracted, efficiency-corrected Dalitz plot distribution of \LbToKSppi decays for \DD and \LL \KS categories combined.
  Some bins have negative entries (consistent with zero) and appear empty.}
\label{fig:dalitz_plot}
\end{figure}

The Dalitz plot distribution of \LbToKSppi decays, shown in Fig.~\ref{fig:dalitz_plot}, is obtained using the \sPlot\ technique and applying event-by-event efficiency corrections based on the position of the decay in the square Dalitz plot.
A structure at low $\proton\pim$ invariant mass, which may originate from excited nucleon states, is apparent but there are no clear structures in the other two invariant mass combinations.

\begin{table}[htb]
\caption{\small
  Fitted yields and efficiency for each channel, separated by \KS type.
  Yields are given with both statistical and systematic uncertainties, whereas for the efficiencies only the uncertainties due to the limited Monte Carlo sample sizes are given.
  The three rows for the \BdToKSpipi decay correspond to the different BDT selections for charmless signal modes and the channels containing \Lc or \Dsm hadrons.
}
\label{tab:fit_results}
\centering
\resizebox{\textwidth}{!}{
\begin{tabular}{lcc|cc}
\hline 
Mode                       &          \multicolumn{2}{c}{\DD}         &          \multicolumn{2}{c}{\LL}        \\
                           &        Yield           &  Efficiency ($\times 10^{-4}$) &        Yield       &  Efficiency ($\times 10^{-4}$)  \\
\hline
\LbToKSppi                 & $\phantom{1}106.1\pm21.5\pm\phantom{1}3.7$   & $5.40\pm0.12$                       & $\phantom{1}90.9\pm 14.6\pm\phantom{1}1.0$  & $2.26\pm0.06$ \\
\LbToKSpK                  & $\phantom{11}{11.5}\pm10.7\pm\phantom{1}1.2$    & $5.34\pm0.11$                       & $\phantom{1}19.6\pm \phantom{1}8.5 \pm\phantom{1}0.8$  & $2.87\pm0.07$ \\
\XibToKSppi                & $ \phantom{111}{5.3} \pm 15.7 \pm \phantom{1}0.7$     & $5.35\pm0.10$                       & $\phantom{11}6.4 \pm \phantom{1}8.5 \pm\phantom{1}0.5$  & $2.67\pm0.07$ \\
\XibToKSpK                 & $\phantom{11}10.5\pm\phantom{1}8.8\pm\phantom{1}0.5$     & $6.12\pm0.10$                       & $\phantom{11}6.3 \pm \phantom{1}5.6 \pm\phantom{1}0.4$  & $2.91\pm0.07$ \\
\hline
\LbToLcpiTopKS             & $1391.6\pm39.6\pm24.8$ & $4.85\pm0.09$                       & $536.8\pm24.6\pm\phantom{1}3.5$  & $1.71\pm0.05$ \\
\LbToLcKTopKS              & $\phantom{11}{70.0}\pm10.3\pm\phantom{1}3.3$    & $4.69\pm0.07$                       & $\phantom{1}37.4 \pm \phantom{1}7.1\pm\phantom{1}2.7$  & $1.66\pm0.03$ \\
\LbToDsp                   & $\phantom{111}{6.3}\pm \phantom{1}5.1\pm \phantom{1}0.6$     & $2.69\pm0.05$                       & $\phantom{11}6.5 \pm \phantom{1}3.7 \pm\phantom{1}0.2$  & $0.89\pm0.03$ \\
\hline
\BdToKSpipi $(\KS\proton h)$ & $\phantom{1}{913.5}\pm45.0\pm12.2$   & $5.57\pm0.09$       & $495.7\pm31.8\pm\phantom{1}{7.5}$  & $2.86\pm0.06$ \\ 
\BdToKSpipi $(\Lc h)$      & $1163.8\pm60.7\pm18.8$                 & $7.38\pm0.11$       & $589.0\pm33.3\pm17.3$              & $3.27\pm0.06$ \\
\BdToKSpipi $(\Dsm p)$     & $1317.8\pm77.1\pm25.7$                 & $7.76\pm0.11$       & $614.1\pm38.3\pm14.8$              & $3.47\pm0.07$ \\
\hline       
\end{tabular} 
}
\end{table} 

\clearpage

\section{Systematic uncertainties}
\label{sec:syst}

The choice of normalisation channel is designed to minimise systematic uncertainties in the branching fraction determination.
Since no $b$~baryon decay has been previously measured with sufficient precision to serve as a normalisation channel, the \BdToKSpipi channel is used.
The remaining systematic uncertainties are summarised in Table~\ref{tab:syst} separately for each signal mode and \KS type.

\begin{table}[b]
  \caption{\small 
    Relative systematic uncertainties on the branching fraction ratios ($\%$) with respect to \BdToKSpipi decays.
    The total is obtained from the sum in quadrature of all contributions except that from knowledge of the fragmentation fractions.}
\label{tab:syst}
\centering
\resizebox{\textwidth}{!}{
\begin{tabular}{ c | c c c c c c |c | c}   
\hline
\DD                                & Simulation     & $\Delta_{\rm PHSP}$ & PID              & Fit model        & Fit bias         & Vetoes           & Total            & $f_{\Lb}/f_{\dquark}$ \\
\hline

${\cal B}(\LbToKSppi)$     	   & $\phantom{1}{6}$ & $\phantom{1}{4}$  & $\phantom{1}{6}$ & $\phantom{1}{1}$ & $<$$1$           & $\phantom{<}{3}$ & $10$             & $27$ \\
${\cal B}(\LbToKSpK)$      	   & $\phantom{1}{6}$ & $58$              & $\phantom{1}{2}$ & $\phantom{1}{8}$ & $\phantom{<}{4}$ & $\phantom{<}{4}$ & $59$             & $27$ \\
${\cal B}(\XibToKSppi)$     	   & $\phantom{1}{4}$ & $64$              & $\phantom{1}{6}$ & $12$             & $\phantom{<}{7}$ & --               & $66$             &  --  \\  
${\cal B}(\XibToKSpK)$      	   & $\phantom{1}{4}$ & $47$              & $\phantom{1}{2}$ & $\phantom{1}{4}$ & $\phantom{<}{3}$ & --               & $47$             &  --  \\   
${\cal B}(\LbToLcpiTopKS)$     	   & $\phantom{1}{5}$ & --                & $\phantom{1}{6}$ & $\phantom{1}{2}$ & $<$$1$           & $<$$1$           & $\phantom{1}{8}$ & $27$ \\
${\cal B}(\LbToLcKTopKS)$      	   & $\phantom{1}{5}$ & --                & $\phantom{1}{4}$ & $\phantom{1}{5}$ & $<$$1$           & $\phantom{<}{1}$ & $\phantom{1}{8}$ & $27$ \\
${\cal B}(\LbToDspToKsK)$  	   & $\phantom{1}{6}$ & --                & $\phantom{1}{6}$ & $\phantom{1}{7}$ & $\phantom{<}{6}$ & --               & $12$             & $27$ \\
\hline
\LL			           &                  &                  &                  &                  &                  &		     &                  &                   \\
\hline
${\cal B}(\LbToKSppi)$             & $\phantom{1}{6}$ & $\phantom{1}{3}$ & $\phantom{1}{4}$ & $\phantom{1}{2}$ & $\phantom{<}{1}$ & $<$$1$           & $\phantom{1}{8}$ & $27$ \\
${\cal B}(\LbToKSpK)$              & $\phantom{1}{6}$ & $42$             & $\phantom{1}{4}$ & $\phantom{1}{4}$ & $\phantom{<}{1}$ & $\phantom{<}{1}$ & $43$             & $27$ \\
${\cal B}(\XibToKSppi)$            & $\phantom{1}{5}$ & $47$             & $\phantom{1}{5}$ & $\phantom{1}{8}$ & $\phantom{<}{2}$ & --               & $49$             & --   \\    
${\cal B}(\XibToKSpK)$             & $\phantom{1}{5}$ & $37$             & $\phantom{1}{5}$ & $\phantom{1}{6}$ & $\phantom{<}{4}$ & --               & $39$             & --   \\   
${\cal B}(\LbToLcpiTopKS)$         & $\phantom{1}{6}$ & --               & $\phantom{1}{4}$ & $\phantom{1}{3}$ & $<$$1$           & $<$$1$           & $\phantom{1}{8}$ & $27$ \\
${\cal B}(\LbToLcKTopKS)$          & $\phantom{1}{5}$ & --               & $\phantom{1}{6}$ & $\phantom{1}{8}$ & $\phantom{<}{1}$ & $<$$1$           & $11$             & $27$ \\
${\cal B}(\LbToDspToKsK)$          & $\phantom{1}{6}$ & --               & $\phantom{1}{8}$ & $\phantom{1}{4}$ & $\phantom{<}{2}$ & --               & $11$             & $27$ \\ 
\hline
\end{tabular}
}
\end{table}

The efficiency determination procedures rely on the accuracy of the simulation.
Uncertainties on the efficiencies arise due to the limited size of the simulation samples, differences between data and the simulation and, for the three-body modes, the variation of the efficiency over the phase-space.

The selection algorithms exploit the difference between signal and background in several variables.
For the \pt and decay length variables, the distributions in data and simulation are known to differ, which can lead to a bias in the estimated efficiency.
The \pt distribution for \LbToLcpi decays in data is obtained with the \sPlot\ technique, and compared to that in the simulation.  
The corresponding possible bias in the efficiency is assigned as systematic uncertainty to each decay.
The value of the \Lb lifetime used in the simulation differs from the most recent measurement~\cite{LHCb-PAPER-2013-032}.
A similar reweighting of the efficiency as done for the \pt distribution results in an estimate of the associated systematic uncertainty for the \Lb modes. 
The \Xib lifetime is not yet measured, and no uncertainty is assigned to the value used in the simulation ($1.42 \ps$) -- unless the true lifetime is dramatically different from this value, the corresponding bias will in any case be negligible compared to other uncertainties.
The uncertainties due to simulation, including also the small effect of limited simulation samples sizes, are combined in quadrature and listed as a single contribution in Table~\ref{tab:syst}.

For modes without significant signals, the effect of efficiency variation across the phase-space (labelled $\Delta_{\rm PHSP}$ in Table~\ref{tab:syst}) is evaluated from the spread of the per-bin efficiency after dividing the square Dalitz plot in a coarse binning scheme.
The large systematic uncertainties reflect the unknown distribution of
signal events across the phase-space and the large efficiency variation.
Conversely, the uncertainties on the normalisation and \LbToKSppi channels are estimated by varying the square Dalitz plot binning scheme. 
For the \BdToKSpipi mode the variation is found to be negligible.
This source of uncertainty does not affect channels with intermediate charmed states, which have known distributions in the phase-space.

The particle identification efficiency and the contamination effects from signal cross-feed contributions are determined with a data-driven method as described in Sec.~\ref{sec:sel}.
In order to estimate possible systematic uncertainties inherent to this procedure, the method is re-evaluated with simulated samples of the control channels.
These average efficiencies are compared to the efficiencies determined from the calibration samples and the differences are taken as estimates of the corresponding systematic uncertainty.
The limited sizes of samples used in the PID calibration also contribute to the systematic uncertainty.

Alternative parametrisations are considered in order to verify the accuracy of the fit model and to assign a systematic uncertainty.
The PDFs of the signal and normalisation channel are replaced respectively with a double CB and the sum of a Gaussian and a bifurcated Gaussian function, 
while the background model is changed to a second-order polynomial function.
The systematic uncertainties are determined from pseudo-experiments, which are fitted with both nominal and alternative models. 
Pseudo-experiments are also used to investigate possible biases induced by the fit model; no significant biases are found, and uncertainties are assigned according to the size of the ensemble.
Finally, the effects of the vetoes applied to remove charmed intermediate states are investigated by studying the variation in the result with different choices of requirements.
The total systematic uncertainty is determined as the sum in quadrature of all contributions. 

The fragmentation fraction of \Lb baryons ($f_{\Lb}$) with respect to those of \Bu and \Bd mesons ($f_u$ and $f_d$, respectively) has been measured by LHCb~\cite{LHCb-PAPER-2011-018} to be 
\begin{equation}
  \label{eq:ff}
  f_{\Lb}/(f_u +f_d) =  (0.404 \pm 0.110)\times [1 - (0.031 \pm 0.005)\times\pt(\!\gevc)]\,,
\end{equation} 
where the statistical, systematic and ${\cal{B}} (\Lc \rightarrow p \Km \pip)$ uncertainties are summed in quadrature,
and the linear dependence is found to apply up to $\pt = 14 \gevc$.
In the case of \Xib baryons, there is no measurement of the fragmentation fraction, and therefore the results quoted include this factor.

The \pt dependence of the fragmentation fraction ratio given in Eq.~(\ref{eq:ff}) is obtained using semileptonic decays, and therefore is given in terms of the combined \pt of the charmed hadron and the muon in the final state.
A correction due to the undetected neutrino is obtained from simulation, so that the appropriate fragmentation fraction ratio corresponding to the mean \pt for each signal mode can be determined ($f_u = f_d$ is assumed)~\cite{LHCb-PAPER-2014-004}.
For channels with significant signal the mean \pt is determined from data with the \sPlot\ technique; otherwise the value from reconstructed simulated events is used. 
Systematic uncertainties arise due to the parametrisation of $f_{\Lb}/f_{d}$ versus \pt and possible inaccuracy in the mean \pt determination.
This results in a fragmentation fraction of $f_{\Lb}/f_{d} = 0.623 \pm 0.030$, $0.590 \pm 0.031$, $0.630 \pm 0.030$, $0.628 \pm 0.030$ and $0.616 \pm 0.030$ for \LbToKSppi, \LbToKSpK, \LbToLcpiTopKS, \LbToLcKTopKS and \LbToDsp decays, respectively.
The large uncertainty due to ${\cal{B}} (\Lc \rightarrow p \Km \pip)$ is not included in these values, but is accounted for separately.

\section{Branching fraction results}
\label{sec:results}

The relative branching fractions are determined according to
\begin{equation}\label{eq:bf_master}
\frac{{\cal{B}}(\LbXibToKSph)}{{\cal{B}}(\BdToKSpipi)} =  
\frac{\epsilon^{\rm sel}_{\Bd\to\KSpipi}}{\epsilon^{\rm sel}_{\LbXib\to\KSph}}
\times \frac{\epsilon^{\rm PID}_{\Bd\to\KSpipi}}{\epsilon^{\rm PID}_{\LbXib\to\KSph}}
\times \frac{N_{\LbXib\to\KSph} }{ N_{\Bd\to\KSpipi}}  
\times \frac{f_{\dquark}}{f_{{\Lb(\Xib)}}},
\end{equation}
where $\epsilon^{\rm sel}$ and $\epsilon^{\rm PID}$ are respectively the selection efficiency (which includes acceptance, reconstruction, offline selection and trigger components) and the particle identification efficiency, $N$ is the signal yield and $f$ is the fragmentation fraction.
Each of these factors is determined separately for each decay and \KS category.
Each pair of results, for \DD and \LL \KS types, is combined in a weighted average, where correlations in the systematic uncertainties are taken into account.
For each mode, the results in the two \KS categories agree within two standard deviations.
For modes with significance below $3\,\sigma$, upper limits are placed at both $90\,\%$ and $95\,\%$ confidence level (CL) by integrating the likelihood multiplied by a Bayesian prior that is uniform in the region of positive branching fraction. 
The following relative branching fraction measurements and limits are obtained
\begin{alignat*}{4}
\frac{{\cal{B}}(\LbToKSppi)}{{\cal{B}}(\BToKSpipi)}     &\,=\,&  &\,0.25  \,\pm 0.04\,\stat  \,\pm\, 0.02\,\syst   \,\pm\, 0.07\,(f_{\Lb}/f_{d})\,, \nonumber \\[2pt]
\frac{{\cal{B}}(\LbToKSpK)}{{\cal{B}}(\BToKSpipi)}      &\,=\,&  &\,0.04  \,\pm 0.02\,\stat  \,\pm\, 0.02\,\syst   \,\pm\, 0.01\,(f_{\Lb}/f_{d})\,, \nonumber \\[-2pt]
                                                        &\,<\,&  &\,0.07~(0.08)~~{\rm at}~~90\,\%\ (95\,\%)~{\rm CL}\,, \nonumber \\[2pt]
f_{\Xib}/f_{d}  \times \frac{{\cal{B}}(\XibToKSppi)}{{\cal{B}}(\BToKSpipi)}  &\,=\,&  &\,0.011  \,\pm 0.015\,\stat  \,\pm\, 0.005\,\syst \,, \nonumber \\[-2pt]
                                                        &\,<\,&  &\,0.03~(0.04)~~{\rm at}~~90\,\%\ (95\,\%)~{\rm CL}\,, \nonumber \\[2pt]
f_{\Xib}/f_{d}  \times \frac{{\cal{B}}(\XibToKSpK)}{{\cal{B}}(\BToKSpipi)}   &\,=\,&  &\,0.012  \,\pm 0.007\,\stat  \,\pm\, 0.004\,\syst \,, \nonumber \\[-2pt]
                                                        &\,<\,&  &\,0.02~(0.03)~~{\rm at}~~90\,\%\ (95\,\%)~{\rm CL}\,, \nonumber \\[2pt]
\frac{{\cal{B}}(\LbToLcpiTopKS)}{{\cal{B}}(\BToKSpipi)} &\,=\,&  &\,2.83  \,\pm 0.13\,\stat  \,\pm\, 0.16\,\syst   \,\pm\, 0.77\,(f_{\Lb}/f_{d})\,, \nonumber  \\[2pt] 
\frac{{\cal{B}}(\LbToLcKTopKS)}{{\cal{B}}(\BToKSpipi)}  &\,=\,&  &\,0.17  \,\pm 0.02\,\stat  \,\pm\, 0.01\,\syst   \,\pm\, 0.05\,(f_{\Lb}/f_{d})\,, \nonumber   \\[2pt]
\frac{{\cal{B}}(\LbToDspToKsK)}{{\cal{B}}(\BToKSpipi)}  &\,=\,&  &\,0.040  \,\pm 0.021\,\stat  \,\pm\, 0.003\,\syst   \,\pm\, 0.011\,(f_{\Lb}/f_{d})\,, \nonumber \\[-2pt]
                                                        &\,<\,&  &\,0.07~(0.08)~~{\rm at}~~90\,\%\ (95\,\%)~{\rm CL}\,. \nonumber
\end{alignat*}
The relative branching fraction of \LbToLcK and \LbToLcpi decays is
\begin{equation*}
  \frac{{\cal{B}}(\LbToLcK)}{{\cal{B}}(\LbToLcpi)} = 0.059 \pm 0.007\,\stat\, \pm 0.004\,\syst\,.
\end{equation*}
This result is in agreement with a recent, more precise measurement~\cite{LHCb-PAPER-2013-056}, from which it is independent, up to a negligible correlation in the systematic uncertainty due to particle identification efficiencies.
The absolute branching fractions are calculated using the measured branching fraction of the normalisation channel 
${\cal{B}}(\BdToKzpipi) = ( 4.96 \pm 0.20 ) \times 10^{-5}$~\cite{PDG2012}.
The results are expressed in terms of final states containing either $\Kz$ or $\Kzb$ mesons, according to the expectation for each decay, 
\begin{alignat*}{5}
{\cal{B}}(\LbToKzbppi)     &= (1.26 \pm 0.19 \pm 0.09 \pm 0.34 \pm 0.05) \times 10^{-5} \,, \nonumber \\[1pt]
{\cal{B}}(\LbToKzpK)       &= (1.8 \pm 1.2 \pm 0.8 \pm 0.5 \pm 0.1) \times 10^{-6} \,, \nonumber \\[-1pt]
&<  3.5~(4.0)  \times 10^{-6}~~{\rm at}~~90\,\%\ (95\,\%)~{\rm CL}\,, \nonumber  \\[1pt]
f_{\Xib}/f_{d}  \times {\cal{B}}(\XibToKzbppi) &= (0.6 \pm 0.7 \pm 0.2) \times 10^{-6} \nonumber \\[-1pt]
&<  1.6~(1.8)  \times 10^{-6}~~{\rm at}~~90\,\%\ (95\,\%)~{\rm CL}\,, \nonumber \\[1pt]
f_{\Xib}/f_{d}  \times {\cal{B}}(\XibToKzbpK)  &= (0.6 \pm 0.4 \pm 0.2) \times 10^{-6}\,, \nonumber \\[-1pt]
&<  1.1~(1.2)  \times 10^{-6}~~{\rm at}~~90\,\%\ (95\,\%)~{\rm CL}\,, \nonumber \\[1pt]
{\cal{B}}(\LbToLcpiTopKz)  &= (1.40  \,\pm 0.07\, \pm 0.08\, \pm 0.38\, \pm 0.06) \times 10^{-4} \,, \nonumber \\[1pt]
{\cal{B}}(\LbToLcKTopKz)   &= (0.83  \,\pm 0.10\, \pm 0.06\, \pm 0.23\, \pm 0.03) \times 10^{-5} \,, \nonumber \\[1pt]
{\cal{B}}(\LbToDspToKzK)   &= (2.0 \,\pm 1.1\, \pm 0.2\, \pm 0.5\, \pm 0.1) \times 10^{-6} \,,\nonumber \\[-1pt]
&<  3.5~(3.9) \times 10^{-6}~~{\rm at}~~90\,\%\ (95\,\%)~{\rm CL}\,, \nonumber
\end{alignat*}
where, for the $\Lb$ decays, the first uncertainty is statistical, the second systematic, the third from $f_{\Lb}/f_{d}$ and the last due to the uncertainty on ${\cal{B}}(\BdToKzpipi)$. 
For the $\Xib$ decays the unknown ratio of fragmentation fractions $f_{\Xib}/f_{d}$ is factored out, and the normalisation channel uncertainty is negligible and is therefore not included.

The $\Lb\to\Lc h^{-}$ absolute branching fractions can be determined more precisely 
than the product branching fractions with $\Lc \rightarrow p \Kzb$, since 
${\cal{B}} (\Lc \rightarrow p \Kzb)/{\cal{B}} (\Lc \rightarrow p \Km \pip)$ is known to better precision~\cite{PDG2012} than
the absolute value of ${\cal{B}} (\Lc \rightarrow p \Km \pip)$ that dominates the uncertainty on $f_{\Lb}/f_{d}$. 
Dividing the product branching fractions quoted above by ${\cal{B}} (\Lc \rightarrow p \Km \pip)$ and by the ratio of \Lc branching fractions gives 
\begin{alignat*}{5}
{\cal{B}}(\Lb \to \Lc\pim) &= (5.97  \,\pm 0.28\, \pm 0.34\, \pm 0.70\, \pm 0.24\,) \times 10^{-3} \,, \nonumber \\
{\cal{B}}(\Lb \to \Lc\Km)  &= (3.55  \,\pm 0.44\, \pm 0.24\, \pm 0.41\, \pm 0.14\,) \times 10^{-4} \,. \nonumber
\end{alignat*}
Similarly, the known value of ${\cal B}(\Dsm\to\KS\Km)$~\cite{PDG2012} can be used to obtain
\begin{alignat*}{5}
{\cal{B}}(\LbToDsp)   &= ( 2.7 \,\pm 1.4 \,\pm 0.2 \,\pm 0.7 \,\pm 0.1 \,\pm 0.1)\times 10^{-4} \,,\nonumber \\
&<  4.8~(5.3) \times 10^{-4}~~{\rm at}~~90\,\%\ (95\,\%)~{\rm CL}\,, \nonumber
\end{alignat*}
where the last uncertainty is due to the uncertainty on ${\cal B}(\Dsm\to\KS\Km)$. 

\section{\boldmath Direct \CP asymmetry}
\label{sec:results-acp}

The significant signal observed for the \LbToKSppi channel allows a measurement of its \CP asymmetry integrated over phase-space.
The simultaneous extended maximum likelihood fit is modified to allow the determination of the raw asymmetry, defined as 
\begin{equation}
  \label{eq:ACP_master}
  {\cal{A}}^{\rm RAW}_{\CP} = \frac{N_{\bar{f}} - N_{f}}{N_{\bar{f}} + N_{f}}\,,
\end{equation}
where $N_{\bar{f}/f}$ is the observed yield for \Lb/\Lbbar decays. 
To obtain the physical \CP asymmetry, this has to be corrected for small detection (${\cal{A}}_{\rm D}$) and production (${\cal{A}}_{\rm P}$) asymmetries, 
${\cal{A}}_{\CP} = {\cal{A}}^{\rm RAW}_{\CP} - {\cal{A}}_{\rm P} - {\cal{A}}_{\rm D}$.
This can be conveniently achieved with \LbToLcpiTopKS decays, which share the same final state as the mode of interest, and have negligible expected \CP violation.

The measured inclusive raw asymmetry for \LbToLcpiTopKS decays is found to be ${\cal{A}}^{\rm RAW}_{\CP}=  -0.047 \pm 0.027$, indicating that the combined detection and production asymmetry is at the few percent level.
The fitted raw asymmetry for \LbToKSppi decays is 
${\cal{A}}^{\rm RAW}_{\CP}=  0.17 \pm 0.13$, 
where the uncertainty is statistical only. 
The raw asymmetry for each of the background components is found to be consistent with zero, as expected.

Several sources of systematic uncertainties are considered.
The uncertainty on ${\cal{A}}_{\rm P} + {\cal{A}}_{\rm D}$ comes directly from the result of the fit to \LbToLcpiTopKS decays.
The effect of variations of the detection asymmetry with the decay kinematics, which can be slightly different for reconstructed \LbToKSppi and \LbToLcpiTopKS decays, is negligible.
The possible variation of the \CP asymmetry across the phase-space of the \LbToKSppi decay, and the non-uniform efficiency results in a systematic uncertainty that is evaluated by weighting events using the \sPlot\ technique and obtaining an efficiency-corrected value of ${\cal{A}}^{\rm RAW}_{\CP}$.
The $0.003$ difference with respect to the nominal value is assigned as uncertainty.
Effects related to the choices of signal and background models, and possible intrinsic fit biases, are evaluated in a similar way as for the branching fraction measurements, leading to an uncertainty of $0.001$.
These uncertainties are summed in quadrature to yield the total systematic uncertainty.

The phase-space integrated \CP asymmetry is found to be 
\begin{equation}
  \label{eq:ACP_Lb2KSppi_final}
  \ACP (\LbToKSppi) = 0.22 \pm 0.13 \,\stat \pm 0.03 \,\syst\,,
  \nonumber
\end{equation}
which is consistent with zero.

\section{Conclusions}
\label{sec:conclusions}

Using a data sample collected by the LHCb experiment corresponding to an integrated luminosity of $1.0 \invfb$ of $pp$ collisions at $\sqrt{s}=7\tev$, searches for the three-body charmless decay modes \LbXibToKSppi and \LbXibToKSpK are performed.
Decays with intermediate charmed hadrons giving the same final state are also investigated.
The decay channel \LbToKSppi is observed for the first time, with a significance of $8.6\,\sigma$, allowing a measurement of its phase-space integrated \CP asymmetry, which shows no significant deviation from zero. 
All presented results, except for those of the branching fractions of $\Lb\to\Lc\pim$ and $\Lb\to\Lc\Km$, are the first to date.
The first observation of a charmless three-body decay of a $b$ baryon opens a new field of possible amplitude analyses and \CP violation measurements that will be of great interest to study with larger data samples.

\section*{Acknowledgements}

\noindent We express our gratitude to our colleagues in the CERN
accelerator departments for the excellent performance of the LHC. We
thank the technical and administrative staff at the LHCb
institutes. We acknowledge support from CERN and from the national
agencies: CAPES, CNPq, FAPERJ and FINEP (Brazil); NSFC (China);
CNRS/IN2P3 and Region Auvergne (France); BMBF, DFG, HGF and MPG
(Germany); SFI (Ireland); INFN (Italy); FOM and NWO (The Netherlands);
SCSR (Poland); MEN/IFA (Romania); MinES, Rosatom, RFBR and NRC
``Kurchatov Institute'' (Russia); MinECo, XuntaGal and GENCAT (Spain);
SNSF and SER (Switzerland); NAS Ukraine (Ukraine); STFC (United
Kingdom); NSF (USA). We also acknowledge the support received from the
ERC under FP7. The Tier1 computing centres are supported by IN2P3
(France), KIT and BMBF (Germany), INFN (Italy), NWO and SURF (The
Netherlands), PIC (Spain), GridPP (United Kingdom).
We are indebted to the communities behind the multiple open source software packages we depend on.
We are also thankful for the computing resources and the access to software R\&D tools provided by Yandex LLC (Russia).

\addcontentsline{toc}{section}{References}
\setboolean{inbibliography}{true}
\bibliographystyle{LHCb}
\bibliography{main,LHCb-PAPER,LHCb-CONF,LHCb-DP}

\end{document}